\documentclass{ws-ijmpe}
\usepackage[super,compress]{cite}
\usepackage{braket}
\begin{document}

\markboth{Kosuke Nomura}
{Microscopic formulation of the interacting boson model for reflection asymmetric nuclei}

\catchline{}{}{}{}{}

\title{Microscopic formulation of the interacting boson model for reflection asymmetric nuclei}

\author{Kosuke Nomura}

\address{Department of Physics, 
University of Zagreb\\
Zagreb, HR-10000, Croatia\\
knomura@phy.hr}

\maketitle

\begin{history}
\received{Day Month Year}
\revised{Day Month Year}
\end{history}

\begin{abstract}
Reflection asymmetric, octupole 
shapes in nuclei are 
a prominent aspect of nuclear 
structure, and have been recurrently studied 
over the decades. 
Recent experiments using radioactive-ion 
beams have provided evidence for stable 
octupole shapes. A variety of 
nuclear models have been employed 
for the related theoretical analyses. 
We review recent studies on the nuclear 
octupole shapes and collective excitations
within the interacting boson model. 
A special focus is placed on the 
microscopic formulation of this model 
by using the mean-field method that is 
based on the framework of 
nuclear density functional theory. 
As an illustrative example, 
a stable octupole deformation, 
and a shape phase transition as
a function of nucleon number 
that involves both quadrupole 
and octupole degrees of freedom 
are shown to occur in light actinides. 
Systematic spectroscopic studies indicate  
enhancement of the octupole collectivity 
in a wide mass region. 
Couplings between the octupole and additional 
degrees of freedom are incorporated in a 
microscopic manner in the boson system, 
and shown to play a crucial role 
in the description of the related intriguing 
nuclear structure phenomena such 
as the shape coexistence. 
\end{abstract}


\ccode{PACS numbers: 21.10.Re, 21.60.Ev, 21.60.Fw, 21.60.Jz}

\section{Introduction}

Reflection asymmetric, octupole deformation 
of the atomic nuclei presents 
a theme of great interest 
in nuclear structure physics 
\cite{butler1996,butler2016}. 
The octupole deformation is expected 
to emerge in those mass regions in 
the chart of nuclides 
in which a coupling occurs between 
the single-particle 
orbitals that differ by 
$\Delta \ell$ = $\Delta j$ = $3\hbar$ 
($\ell$ and $j$ are quantum numbers 
of single-particle state) and are 
opposite in parity. 
Empirical signatures of the octupole deformation 
are low-energy negative-parity states forming 
an alternating parity band with 
the positive-parity yrast states, 
and enhanced electric dipole 
and octupole transitions within the band.

The existence of low-lying negative-parity 
states that are associated with the reflection 
asymmetric deformation was recognized in the 1950's. 
Since then the topic has been recurrently 
pursued. 
Recent experiments using radioactive-ion-beams 
have identified evidence for stable octupole shapes, 
e.g., in light actinides 
\cite{gaffney2013,chishti2020,butler2020a} 
and lanthanides \cite{bucher2016,bucher2017}.
Besides that, theoretical studies 
that are to support and predict 
the occurrence of nuclear octupole states have 
been extensively conducted using a variety of 
nuclear structure models. 
An extensive list of the relevant 
experimental and theoretical studies 
is given in the review articles, e.g., 
Refs.~\refcite{butler1996,butler2016}. 
In addition, the presence of static 
octupole correlations in the nucleus 
enhances an atomic electric dipole moment (EDM). 
The observation of a non-zero EDM would imply the 
violation of the CP symmetry, hence is 
important for exploring new physics beyond 
the standard model of elementary particles 
\cite{engel2013}.

In this paper, 
we review recent 
theoretical investigations on the 
reflection asymmetric nuclear shapes 
and related collective states 
within the interacting boson model (IBM) \cite{IBM}. 
A special focus is placed on the microscopic 
formulation of this model in terms of 
the nuclear mean-field method. 
Within this theoretical scheme 
a stable octupole shape emerges 
in a characteristic set of nuclei in which 
octupole deformation is empirically suggested 
to occur. 
A transition between the quadrupole and 
octupole states as functions 
of the nucleon number is discussed in the 
context of the quantum phase transition 
in nuclear shapes \cite{cejnar2010}. 
Coupling between single-particle and collective 
octupole degrees of freedom is shown to play 
an important role in determining the low-lying 
band structure of the odd-mass nuclei. 
Furthermore the octupole correlations are shown 
to be relevant for understanding the nature 
of the low-energy excited $0^+$ states, 
which are often considered as a signature 
of shape coexistence \cite{wood1992,heyde2011}.

\section{Interacting boson model for reflection asymmetric nuclei}

The IBM, proposed by Arima and Iachello \cite{arima1975,IBM}, 
has been remarkably successful in 
reproducing low-energy quadrupole collective 
states in a large number of medium-heavy 
and heavy nuclei. 
Its basic assumption is 
that the multi-fermion dynamics of 
the nuclear surface deformation is simulated 
in terms of the monopole, $s$, and quadrupole, 
$d$, bosons, which represent, from a microscopic 
point of view \cite{OAIT,OAI,IBM}, collective $S$ 
and $D$ pairs of valence nucleons 
corresponding to spin and parity 
$0^+$ and $2^+$, respectively. 
This corresponds to 
a drastic truncation 
of the full shell model configuration space 
which becomes prohibitively large 
for heavy and open-shell nuclei. 
The model is also intimately related 
to the group theory. 
A prominent feature 
is the emergence of the dynamical symmetry, 
i.e., if the Hamiltonian is written as a specific 
combination of Casimir operators 
then the Hamiltonian is associated with a 
certain intrinsic structure and is exactly solvable. 
The $sd$-IBM, for instance, constitutes 
the bosonic algebra U(6), and from it appear 
three subalgebras U(5), SU(3), 
and O(6), which correspond to vibrational, 
rotational, and $\gamma$-unstable limits of the 
quadrupole mode, respectively.

To study negative-parity states, 
the IBM was extended 
\cite{arima1978su3,engel1985,engel1987,barfield1988}
so that the boson model space should contain, 
in addition to 
the positive-parity $s$ and $d$ bosons, 
the negative-parity bosons such as the 
dipole, $p$, and 
octupole, $f$, bosons representing 
the spin and parity $1^-$ and $3^-$ 
nucleon pairs, respectively. 
The $spdf$- and $sdf$-boson models 
have been employed 
in phenomenological studies 
on the quadrupole-octupole coupled collective 
excitations in realistic cases in 
actinides \cite{zamfir2001}, 
rare-earth \cite{cottle1996} and 
lanthanides \cite{kusnezov1988}. 
From a group theoretical point of view, 
the $sdf$- and $spdf$-boson models constitute, 
respectively, the boson algebras U(13) and U(16). 
The group structures including the dynamical 
symmetries of the quadrupole 
and octupole states have 
been analyzed in this respect 
\cite{engel1985,engel1987,kusnezov1989,kusnezov1990}.

Besides numerous successes in reproducing 
observed collective spectra, 
the IBM should have its microscopic 
basis on nucleonic degrees of freedom, 
and attempts have been made to establish a link 
between the IBM and more microscopic nuclear 
structure models. 
A standard method, referred to as the 
Otsuka-Arima-Iachello (OAI) mapping 
\cite{OAIT,OAI}, is such that a seniority-based 
shell-model state given in terms of the 
$S$, $D$, $\ldots$ nucleon pairs is mapped onto 
the equivalent $s$, $d$, $\ldots$ boson state. 
The number of bosons is set 
equal to that of the valence nucleons. 
The OAI method 
has been shown to be valid in limited realistic 
cases of moderately deformed 
nuclei, i.e., nearly 
spherical and $\gamma$-unstable ones 
\cite{mizusaki1996}.

More recently, making use of the fact 
that the potential energy surface of a given 
nuclear mean-field model can be simulated 
by that of the boson system, 
the IBM Hamiltonian has been shown \cite{nomura2008} 
to be derived in the general situations 
of quadrupole collective mode, 
covering the nearly spherical vibrational [U(5)] 
\cite{nomura2008,nomura2010}, 
strongly deformed rotational [SU(3)] 
\cite{nomura2011rot}, 
$\gamma$-unstable [O(6)] 
\cite{nomura2008,nomura2010}, 
and triaxially deformed states 
\cite{nomura2012tri}. 
The method was extended further to 
address various problems in nuclear structure, 
including the shape coexistence 
\cite{nomura2012sc}, 
octupole deformation and collectivity 
\cite{nomura2013oct,nomura2014,nomura2015}, 
description of odd-mass nuclei 
\cite{nomura2016odd}, 
and nuclear $\beta$ decays 
\cite{nomura2020beta-2,nomura2020beta-1}. 
In what follows, we outline the above-mentioned 
method and discuss its applications 
to those nuclear properties in which octupole 
correlations are expected 
to be relevant.

\begin{figure}[th]
\begin{tabular}{cc}
\includegraphics[width=.5\linewidth]{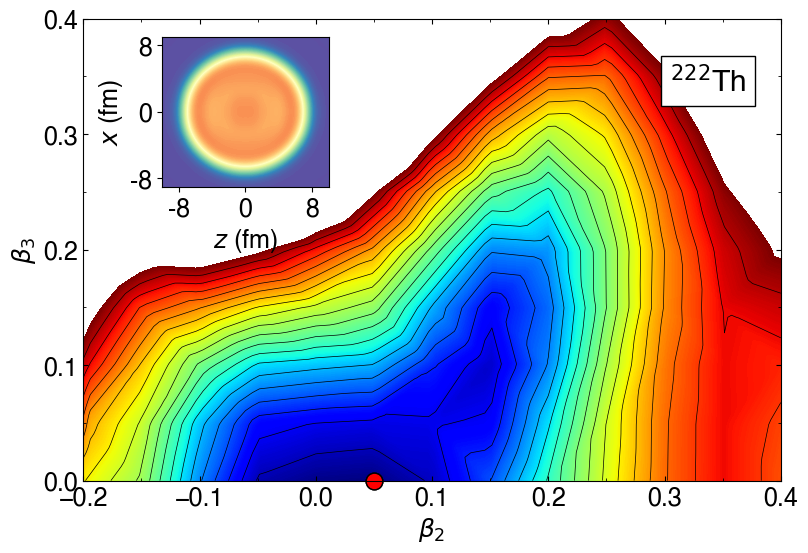} &
\includegraphics[width=.5\linewidth]{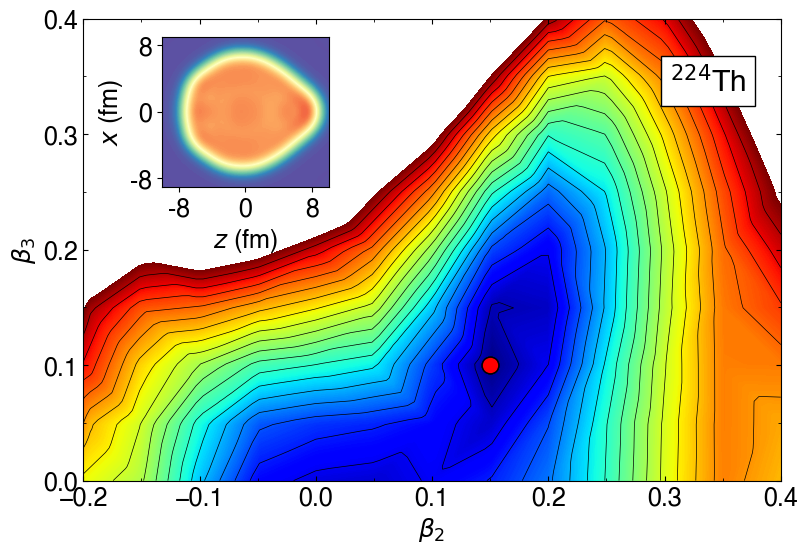} \\
\includegraphics[width=.5\linewidth]{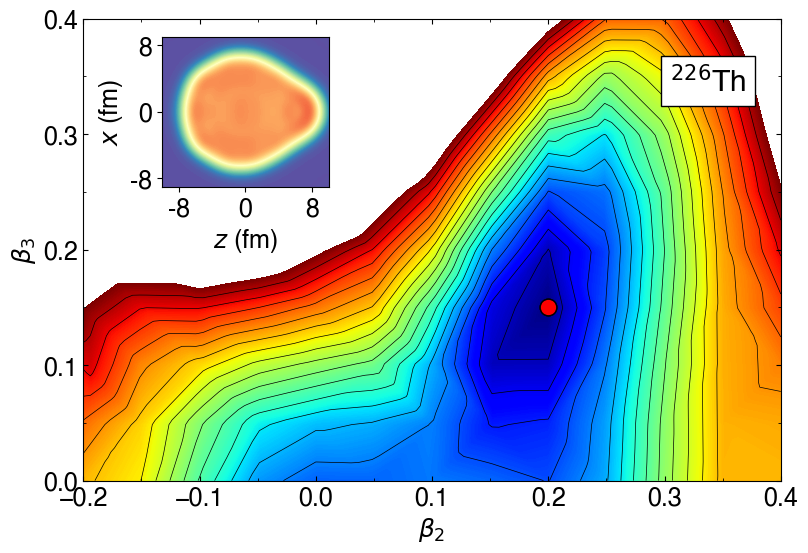} &
\includegraphics[width=.5\linewidth]{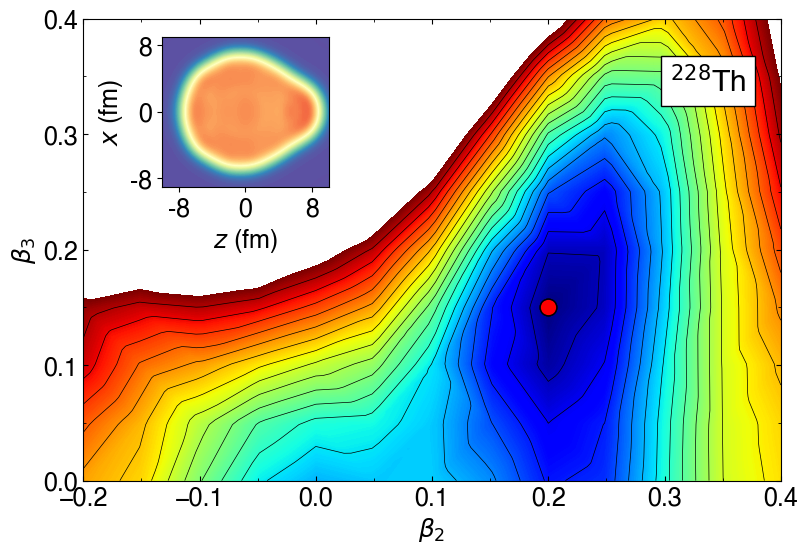} \\
\includegraphics[width=.5\linewidth]{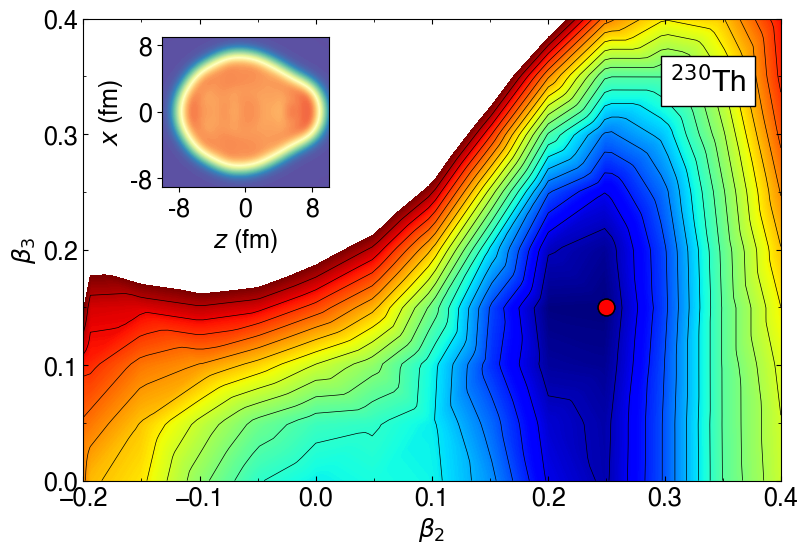} &
\includegraphics[width=.5\linewidth]{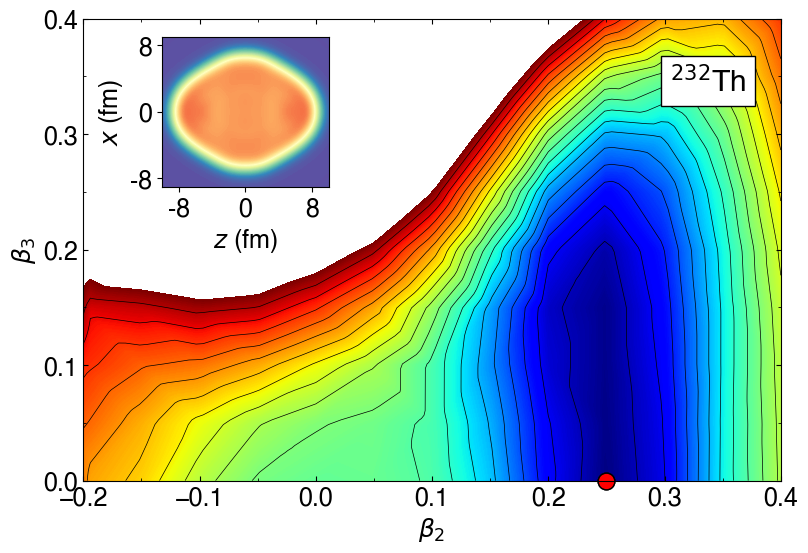}
\end{tabular}
\caption{Potential energy surfaces for $^{222-232}$Th 
in terms of the axially-symmetric 
quadrupole $\beta_{2}$ and octupole 
$\beta_{3}$ deformations, 
calculated within the relativistic Hartree-Bogoliubov 
method using the functional DD-PC1 and 
separable pairing force. 
The energy surface is plotted up to 10 MeV 
from the global minimum, indicated by the solid 
red circle. The energy difference between 
the neighboring contour lines is 0.5 MeV.
The calculated intrinsic nucleon density 
in the $x$-$z$ plane is also shown 
for each nucleus. 
}
\label{fig:pes-th}
\end{figure}

\section{Mean-field derivation of the interacting boson model}

\subsection{Mean-field calculations}

Among contemporary theoretical approaches, 
the nuclear density functional theory (DFT)
\cite{bender2003,vretenar2005,niksic2011,robledo2019,schunck2019} 
allows for a global and accurate prediction of 
bulk nuclear matter and 
intrinsic properties 
of finite nuclei, including 
the mass, radii, deformations, etc., 
and collective excitations 
over the entire region of 
the nuclear chart. 
Its basic implementation 
is in the self-consistent mean-field (SCMF) calculations, 
in which an energy density functional (EDF) 
is constructed as a functional of one-body 
density matrices that correspond to a 
single product state. 
Both nonrelativistic \cite{bender2003,robledo2019} 
and relativistic \cite{vretenar2005,niksic2011} 
EDFs have been successfully 
applied to numerous studies on nuclear 
structure phenomena.

The starting point is a set of 
constrained SCMF calculations \cite{RS} 
for each nucleus 
to obtain the potential energy surface, 
that is, 
the total mean-field energy as 
a function of the relevant shape degrees 
of freedom. 
Here the constraints are on the axially 
symmetric intrinsic quadrupole $Q_{20}$ 
and octupole $Q_{30}$ moments, 
which are related to  
the deformation variables $\beta_{2}$ 
and $\beta_{3}$, respectively. 

As an illustrative example, 
Fig.~\ref{fig:pes-th} shows the mean-field 
potential energy surfaces for the 
axially symmetric even-even nuclei 
$^{222-232}$Th. 
The SCMF calculations 
are performed within the relativistic 
Hartree-Bogoliubov (RHB) method 
\cite{vretenar2005,niksic2011} 
based on the 
density-dependent point-coupling (DD-PC1) 
\cite{DDPC1} interaction and the separable 
pairing force of finite range \cite{tian2009}. 
Variation of the topology of energy surface 
as a function of nucleon number gives a 
qualitative interpretation of nuclear shape 
evolution. 
The global (equilibrium) minimum 
occurs at nearly spherical 
mean-field configuration 
($\beta_2$, $\beta_3$) $\approx$ (0.05, 0.0) 
for $^{222}$Th. 
In the neighboring isotope $^{224}$Th 
a non-zero octupole minimum with 
$\beta_3$ $\approx$ 0.1 is seen. 
The octupole deformation becomes most 
pronounced for $^{226}$Th. 
Note the occurrence of the stable 
octupole minimum is inferred 
by the behaviors of the 
single-particle levels near the Fermi energies; 
At the deformations that correspond to the 
octupole global minimum, level densities of 
the single proton and neutron orbits 
were indeed shown to be lower near the 
Fermi energies \cite{li2016}. 
For heavier nuclei $^{228,230}$Th, 
the energy surface becomes softer along 
the $\beta_3$ direction. 
For $^{232}$Th, the octupole minimum 
disappears, but the potential is 
substantially soft in the $\beta_3$ 
deformation.

Figure~\ref{fig:pes-th} also shows 
total intrinsic nucleon density 
in the $x$-$z$ plane for each nucleus, 
computed within the RHB method 
with the constraints on 
those $\beta_2$ and $\beta_3$ 
deformations corresponding to the global minimum. 
One finds the intrinsic 
nucleon densities resembling 
a completely spherically symmetric shape 
for $^{222}$Th, a reflection 
asymmetric shape for 
$^{224-230}$Th, and a reflection symmetric 
shape for $^{232}$Th.

\subsection{Mapping onto the boson system\label{sec:mapping}}

Let us turn to the boson system. 
We employ the simplest version 
of the $sdf$-boson model, 
where no distinction is made between proton 
and neutron bosons. 
A form of the $sdf$-IBM Hamiltonian 
appropriate for realistic calculations 
\cite{nomura2013oct,nomura2014} reads
\begin{eqnarray}
\label{eq:bham}
\hat H_\text{B}
= \epsilon_d \hat n_d 
+ \epsilon_f \hat n_f 
+ \kappa_2 \hat Q \cdot \hat Q
+ \kappa_2' \hat L_d \cdot \hat L_{d} 
+ \kappa_3 : \hat V_3^\dagger \cdot \hat V_3 :
\; .
\end{eqnarray}
$\hat n_d = d^\dagger \cdot \tilde d$ and 
$\hat n_f = f^\dagger \cdot \tilde f$ are 
number operators for $d$ and $f$ bosons, 
with $\epsilon_d$ and $\epsilon_f$ being 
single-$d$ and -$f$ boson energies 
relative to the $s$ boson one, respectively. 
Note that 
$\tilde d_\mu = (-1)^{-\mu} d_{-\mu}$, 
$\tilde f_\mu = (-1)^{3-\mu} f_{-\mu}$, 
and the dot ($\cdot$) means scalar product. 
The third term in Eq.~(\ref{eq:bham}) 
is the quadrupole-quadrupole 
interaction with the strength 
parameter $\kappa_2$, and 
$\hat Q = s^\dagger \tilde d + d^\dagger s + \chi_d [d^\dagger \times \tilde d]^{(2)} + \chi_f [f^\dagger \times \tilde f]^{(2)}$ 
is the quadrupole operator. 
$\chi_d$ and $\chi_f$  
are dimensionless parameters. 
The fourth term, with the strength 
$\kappa_2'$, is specifically required 
for strongly axially deformed 
nuclei \cite{nomura2011rot}, 
and $\hat L_d$ is the 
angular momentum operator for $d$ bosons 
$\hat L_d = \sqrt{10} [d^\dagger \times \tilde d]^{(1)}$. 
The last term, expressed 
in the normal-ordered form, 
represents the octupole-octupole 
interaction with the strength parameter 
$\kappa_3$, and the operator 
$\hat V_3^\dagger =  s^\dagger \tilde f + \chi_3 [d^\dagger \times \tilde f]^{(3)}$, with $\chi_3$ being 
another dimensionless parameter. 
For details, see Ref.~\refcite{nomura2014}.

A given boson model Hamiltonian can be related to 
a certain geometrical structure by using 
the boson coherent state 
\cite{ginocchio1980,dieperink1980,bohr1980}. 
For the $sdf$-IBM, if 
axial symmetry is assumed 
the coherent state $\ket{\phi}$ 
is expressed as
\begin{eqnarray}
\label{eq:coherent}
 \ket{\phi} = 
(n!)^{-1/2}
(\lambda^\dagger)^n
\ket{0} \; ,
\quad
\lambda^\dagger
=(1 + \tilde\beta_{2}^2 + \tilde\beta_{3}^2)^{-1/2}
(s^\dagger + \tilde\beta_{2} d_0^\dagger 
+ \tilde\beta_{3} f_0^\dagger) \; .
\end{eqnarray}
Here, $n$ denotes the number of bosons, 
and $\ket{0}$ is the boson 
vacuum, i.e., the inert core. 
The amplitudes $\tilde\beta_{2}$ 
and $\tilde\beta_{3}$ are considered 
as the boson analogs of 
$\beta_2$ and $\beta_3$ deformations in the 
geometrical model. 
The energy surface 
of the boson system, denoted by 
$E_\text{IBM}(\tilde\beta_2, \tilde\beta_3)$, 
is obtained by taking an expectation value  
$\braket{\phi|\hat H_\text{B}|\phi}$, 
which also depends on the parameters 
of the Hamiltonian $\hat H_\text{B}$ 
(\ref{eq:bham}). 
To a good approximation, the 
bosonic deformation variables 
($\tilde\beta_2$, $\tilde\beta_3$) are 
connected to the fermionic counterparts 
($\beta_2$, $\beta_3$) through the relations 
\cite{ginocchio1980,nomura2008,nomura2014}
\begin{eqnarray}
\label{eq:cbeta}
 \tilde\beta_2 = c_2 \beta_2 \; , 
\quad 
\tilde\beta_3 = c_3 \beta_3 \; , 
\end{eqnarray}
with $c_2$ and $c_3$ being constants of 
proportionality. Typical values for these 
constants are $\sim$ 5. The fact that the bosonic 
deformations are larger than the fermionic ones 
reflects that the IBM is built on the limited 
configuration (valence) space while the 
SCMF model considers all nucleons 
in the nucleus. 

\begin{figure}[th]
\centerline{\includegraphics[width=\linewidth]{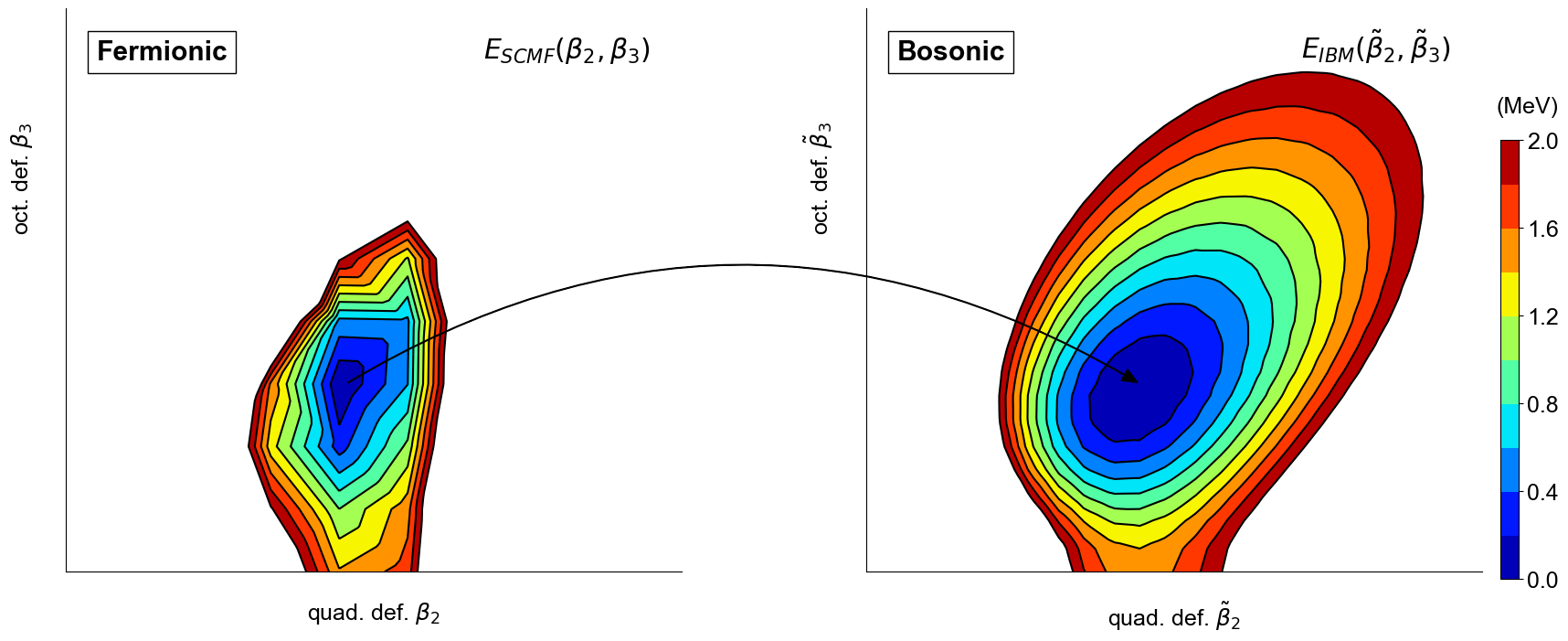}}
\caption{A schematic illustration of the mapping of 
the fermionic potential energy surface 
$E_\text{SCMF}(\beta_{2},\beta_{3})$ 
onto the bosonic counterpart 
$E_\text{IBM}(\tilde\beta_{2},\tilde\beta_{3})$.}
\label{fig:mapping}
\end{figure}

Together with the relations 
in (\ref{eq:cbeta}), 
the SCMF and IBM potential energy surfaces 
are equated to each other: 
\begin{eqnarray}
\label{eq:mapping}
E_\text{SCMF}(\beta_2, \beta_3)
\sim
E_\text{IBM}(\tilde\beta_2, \tilde\beta_3) \; .
\end{eqnarray}
This procedure represents a mapping of the 
fermionic energy surface onto the bosonic one, 
as schematically illustrated in Fig.~\ref{fig:mapping}. 
The parameters of the Hamiltonian 
$\hat H_\text{B}$, and the proportionality 
constants $c_2$ and $c_3$ for the deformation 
variables are determined by the mapping procedure. 
In other words, the IBM parameters are derived 
so that the bosonic energy surface becomes 
similar to the fermionic one as much as possible. 
Note that the relation 
(\ref{eq:mapping}) is an approximate 
equality that should be satisfied 
within a limited range of the ($\beta_2$, $\beta_3$) 
space, that is, in the vicinity of the 
global minimum. 
This is based on the fact that 
the low-energy quadrupole and octupole 
collective states 
are predominantly accounted for by the 
SCMF solutions near the global minimum
\cite{nomura2008,nomura2010,li2016,nomura2021qoch}.

In addition, since the parameter $\kappa_2'$ 
for the rotational term $\hat L_d \cdot \hat L_d$ 
[see Eq.~(\ref{eq:bham}] 
cannot be uniquely determined from the 
comparison of the potential energy surfaces, 
it is derived separately 
\cite{nomura2011rot}, from the comparison of 
the intrinsic moment of inertia in the 
boson system to the fermionic one 
computed in the cranking approximation.

Having determined the parameters 
for $\hat H_\text{B}$ by the mapping procedure, 
the resultant IBM Hamiltonian 
is numerically diagonalized, resulting in the 
energies and wave functions of 
the excited states of both parities. 
It should be noted that no phenomenological 
adjustment is made in this procedure, that is, 
the $sdf$-IBM Hamiltonian is determined only by 
using inputs provided by the microscopic 
calculations within the EDF 
framework, which is also not specifically 
adjusted to the observed octupole states.

At this point, it is worth mentioning that 
the octupole shapes and collective excitations 
have also been extensively studied by 
a number of beyond-mean-field approaches 
within the EDF framework, including 
the symmetry projected generator coordinate method 
(see, e.g., Refs.~\refcite{robledo2019,nomura2020oct}, 
and an extensive list of references are therein)
and the quadrupole-octupole 
collective model 
\cite{li2013,xia2017,nomura2021qoch}.

\subsection{Octupole shape phase transitions in Th isotopes}

\begin{figure}[th]
\centerline{\includegraphics[width=.8\linewidth]{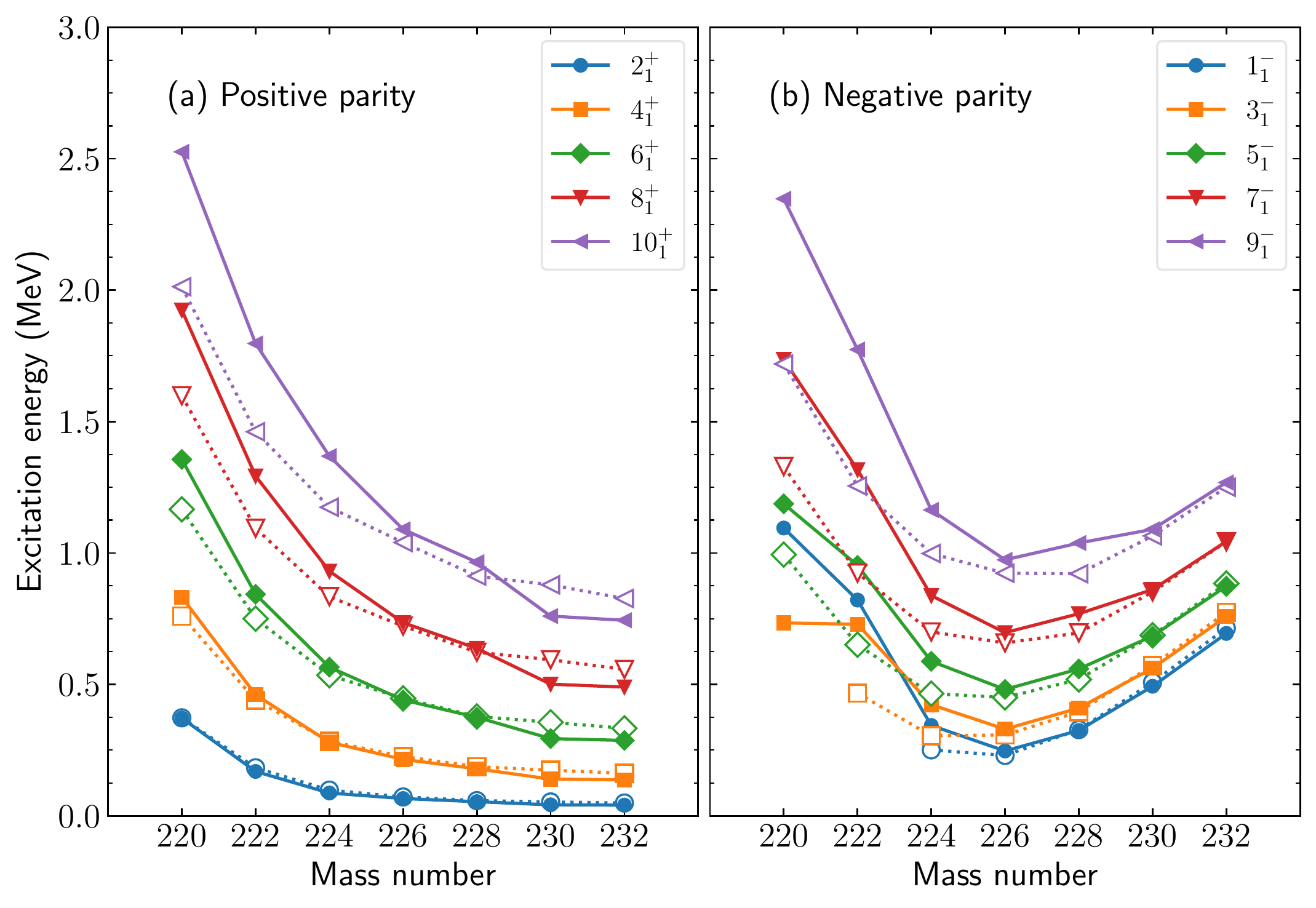}}
\caption{Evolution of the excitation energies 
for the (a) positive- and (b) negative-parity 
yrast states for $^{220-232}$Th. Solid (open) symbols 
represent the calculated (experimental) values, and 
are connected by solid (broken) lines. 
}
\label{fig:level-th}
\end{figure}

Figure~\ref{fig:level-th} shows the calculated
energy spectra for the even-spin positive-parity 
and odd-spin negative-parity yrast states for 
$^{220-232}$Th. Agreement between the calculated 
and experimental \cite{data} energy spectra is good. 
As one sees in Fig.~\ref{fig:level-th}(a), 
the overall decrease in energy of the 
positive-parity levels indicates a transition 
from vibrational to rotational nuclear structures. 
The negative-parity levels, on the other hand, 
exhibit a characteristic parabolic behavior 
with the minimum energy at $^{226}$Th. 
[see Fig.~\ref{fig:level-th}(b)]. 
Around $^{226}$Th, bands with both parities 
are close to each other, and 
appear to form an 
approximate alternating parity band 
as expected for a stable octupole state. 
For $A$ $>$ 226, 
the negative-parity levels keep rising 
with mass number, while the positive-parity 
ones gradually decrease. 
Toward the nucleus 
$^{232}$Th, the positive- and 
negative-parity bands are almost 
decoupled from each other, and 
appear to form separate $\Delta I$ = 2 
rotational bands.

\begin{figure}[th]
\centerline{\includegraphics[width=.8\linewidth]{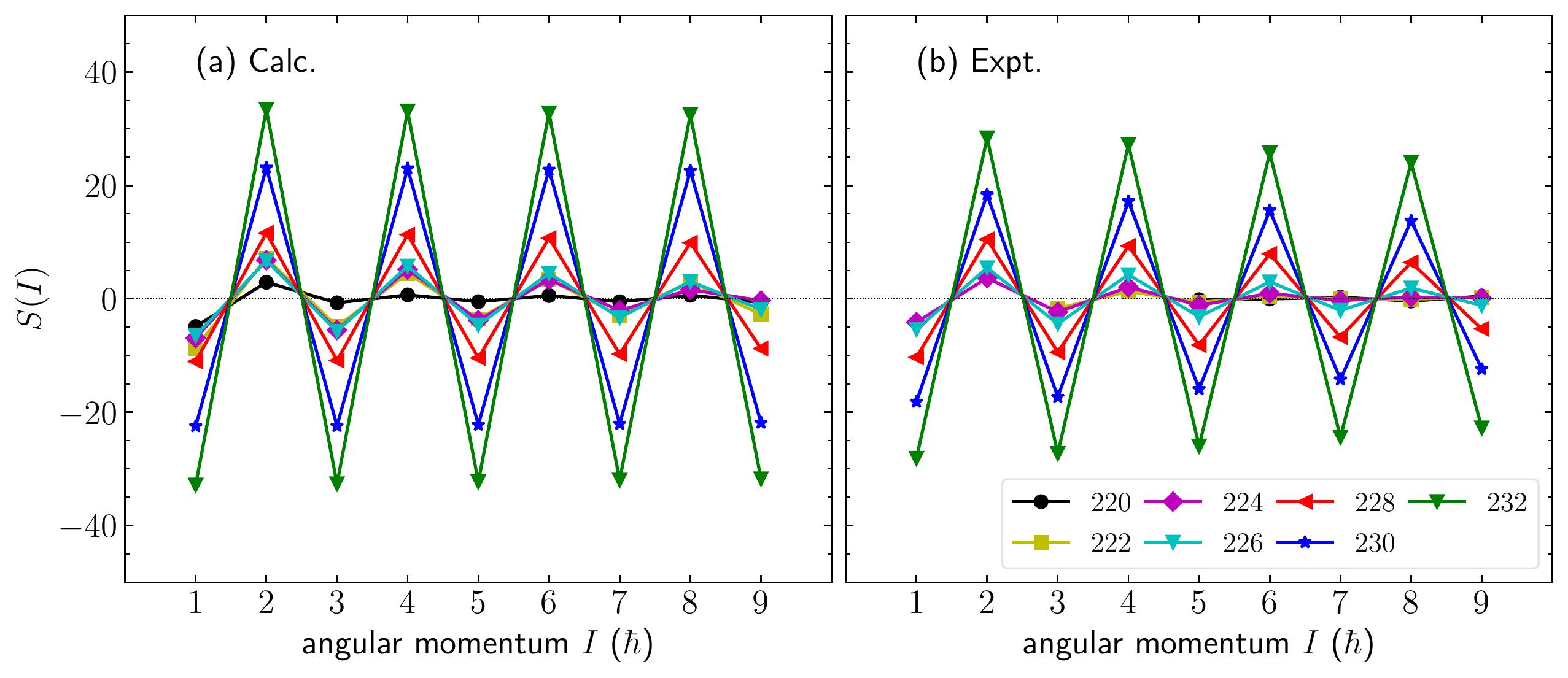}}
\caption{Calculated (a) and experimental 
\cite{data} (b) signature 
splittings $S(I)$ between the even-$I$ positive-parity 
and odd-$I$ negative-parity 
bands for $^{220-230}$Th 
as functions of the angular momentum.}
\label{fig:sgn}
\end{figure}

To identify the occurrence of the 
alternating-parity band structure 
in the Th isotopes, it is convenient to study 
the signature splitting: 
\begin{eqnarray}
 S(I) = \frac{[E(I+1)-E(I)]-[E(I)-E(I-1)]}{E(2^+_1)} \; .
\end{eqnarray}
$E(I)$ stands for the excitation energies 
of the $I$ = $1^-_1$, $2^+_1$, $3^-_1$, $4^+_1$, 
$\ldots$ states. The quantity $S(I)$ 
is sensitive to the splitting between the 
positive- and negative-parity rotational 
bands. For an ideal alternating parity band, 
the positive- and negative-parity bands 
should appear with an equal energy splitting, 
hence $S(I)$ $\approx$ 0. 
Non-zero $S(I)$ value would indicate 
a deviation from the pure alternating parity 
band. 
This situation is illustrated 
in Fig.~\ref{fig:sgn}. 
For $^{220-226}$Th, the $S(I)$ value is nearly 
equal to zero. 
For $^{228-232}$Th, in turn, 
the odd-even spin staggering occurs, 
and the deviation from the limit 
$S(I)$ = 0 becomes even larger 
with mass number. 

The systematic behavior of the calculated 
energy spectra for the yrast bands 
with both parities, point to the occurrence 
of a type of quantum phase transition \cite{cejnar2010} 
along the chain of Th isotopes, 
that is, the structural 
evolution from nearly spherical to stable 
octupole shapes toward $^{226}$Th, and to an 
octupole vibrational state around $^{230,232}$Th 
associated with the octupole-soft potential. 
The behavior of the calculated spectroscopic 
properties reflects, to a large extent, 
the variation of the topology of the 
SCMF energy surface 
in the $\beta_2$-$\beta_3$ plane, as 
was shown in Fig.~\ref{fig:pes-th}. 
In addition, the Th isotopes appears 
to be an ideal case that exhibits 
a stable octupole deformation, as well as 
a shape phase transition that 
involves quadrupole and octupole modes.

\section{Global study}

The octupolarity is 
suggested to be enhanced  
in several different mass regions, 
corresponding to 
the neutron and proton numbers 
($N$, $Z$) $\approx$ (134, 88), (88, 56) 
(56, 56), (56, 34), (34, 34), $\ldots$, 
at which the $\Delta \ell$ = $\Delta j$ = $3\hbar$ 
coupling between the single-particle 
orbitals can occur. 

\begin{figure}[th]
\centerline{\includegraphics[width=.8\linewidth]{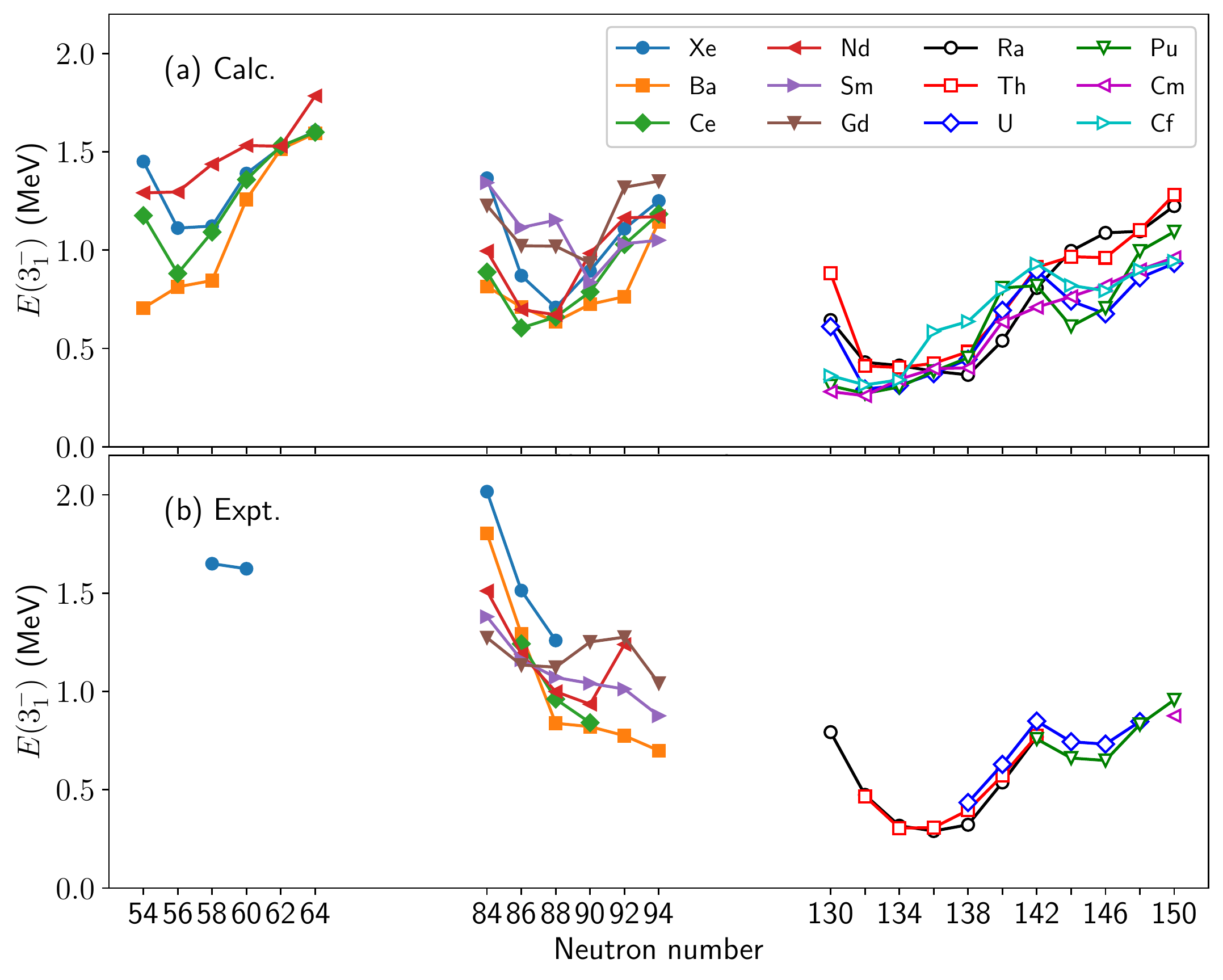}}
\caption{Evolution of (a) 
calculated and (b) experimental 
\cite{data} $3^-_1$ 
excitation energies as functions 
of the neutron number along those isotopic 
chains characteristic of the 
octupole deformation.}
\label{fig:3-}
\end{figure}

Figure~\ref{fig:3-} shows the calculated $3^-_1$ 
excitation energies for the nuclei 
in those characteristic mass regions in which 
octupole deformation is expected to occur. 
The plot has been made as a compilation 
of the results 
\cite{nomura2015,nomura2020oct,nomura2021oct-u,nomura2021oct-ba,nomura2021oct-zn}
obtained from the $sdf$-IBM that is 
based on the Hartree-Fock-Bogoliubov 
(HFB) calculations using the Gogny-D1M 
\cite{D1M} EDF \cite{Gogny}. 
In general, 
the predicted $3^-_1$ energy shows a parabolic 
behavior that is centered around 
$N$ = 56, 88, and 134. 
This tendency is consistent with what is suggested 
experimentally, apart from the Ba and Sm isotopes. 
In many of the nuclei in the light actinides 
close to $N$ = 134, 
the $3^-$ state appears at substantially 
low excitation 
energy $E_x$ $<$ 0.5 MeV. 
The corresponding Gogny-HFB SCMF calculations 
indeed give an octupole-deformed 
($\beta_{3}$ $\neq$ 0)
equilibrium minimum for 
these actinide nuclei 
\cite{nomura2020oct,nomura2021oct-u}. 
One notices in Fig.~\ref{fig:3-} 
another parabolic 
variation of the $3^-$ level around $N$ = 146. 
This local behavior implies the 
effect of dynamical octupole correlations 
that could be important even away 
from the neutron number $N$ = 134. 
The $3^-$ energy is calculated 
to be rather high for the 
neutron-deficient Ba region with 
54 $\leq$ $N$ $\leq$ 64. 
In this region, 
therefore, the octupole correlations 
appear to be relatively weak. 
It is also noted that for the nuclei 
with $N$ $\approx$ $Z$, some additional 
correlations such as the neutron-proton pairs may 
play a role, which are, however, not taken 
into account in the present calculations.

\begin{figure}[th]
\centerline{\includegraphics[width=.8\linewidth]{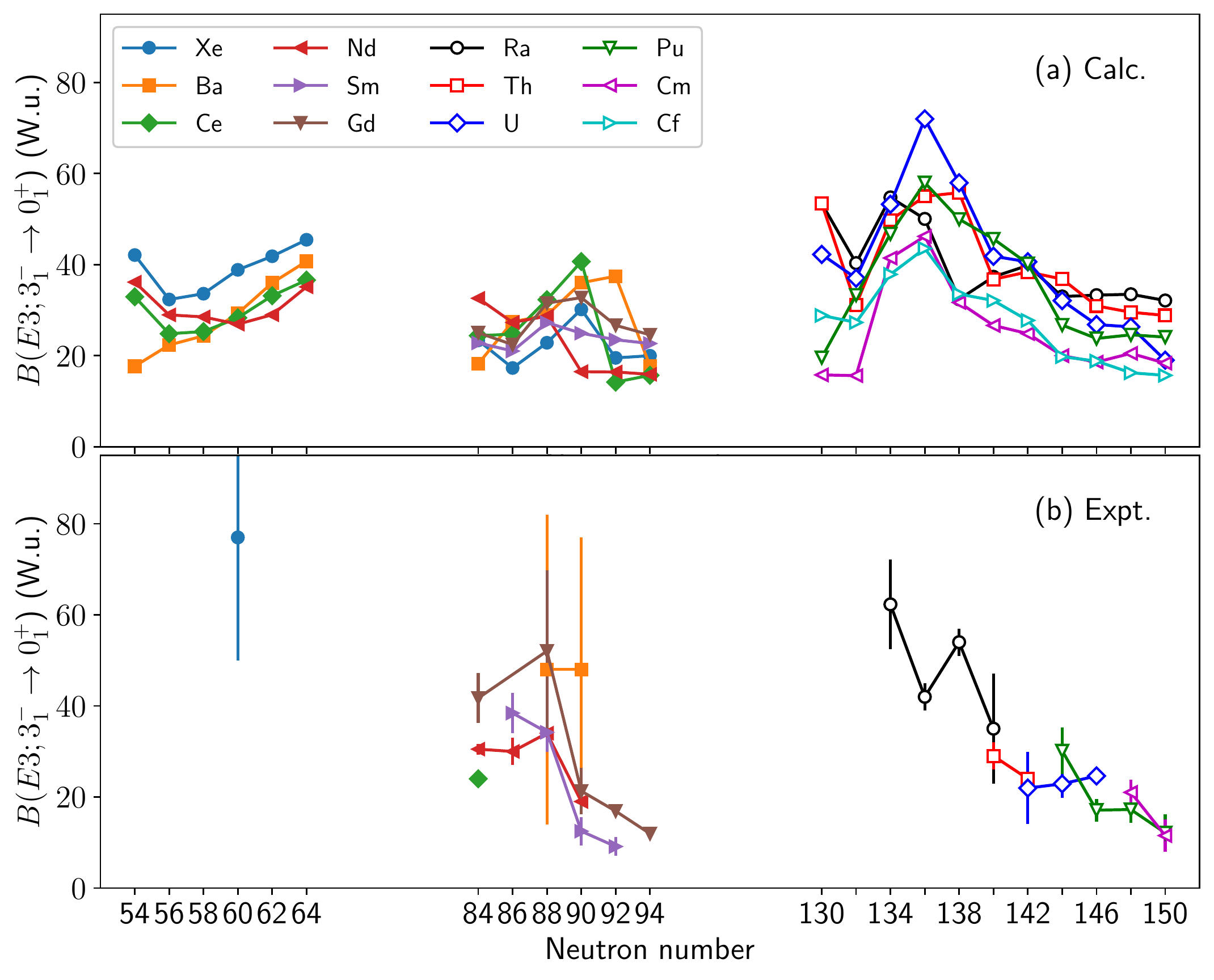}}
\caption{(a) Calculated and (b) experimental 
\cite{data,bucher2016,bucher2017,kibedi2002,DEANGELIS2002} 
$B(E3; 3^-_1 \to 0^+_1)$ transition probabilities 
for the same nuclei as those considered in Fig.~\ref{fig:3-}. 
}
\label{fig:e3}
\end{figure}

The electric octupole ($E3$) 
transition probability 
$B(E3; 3^-_1 \to 0^+_1)$ 
is another indicator of 
the enhanced octupole collectivity. 
Figure~\ref{fig:e3} shows the 
$B(E3; 3^-_1 \to 0^+_1)$ values for the 
same nuclei as those shown in Fig.~\ref{fig:3-}, 
computed by the $sdf$-IBM with the microscopic 
input from the same Gogny-HFB 
calculations 
\cite{nomura2015,nomura2020oct,nomura2021oct-u,nomura2021oct-ba,nomura2021oct-zn}. 
For the $E3$ transition operator, 
the standard boson $E3$ effective charges 
that are often used in realistic 
$sdf$-IBM calculations are chosen 
\cite{nomura2015,nomura2020oct,nomura2021oct-u,nomura2021oct-ba,nomura2021oct-zn}.
The calculated $B(E3; 3^-_1 \to 0^+_1)$ 
rate exhibits an inverse 
parabolic behavior exhibiting a maximum 
value around $N$ = 88 and 134, 
This behavior corroborates the one 
obtained for the $3^-$ energy 
level (see Fig.~\ref{fig:3-}), 
and is consistent with the data.

\section{Coupling with additional degrees of freedom}

\subsection{Particle-boson coupling in odd-mass systems\label{sec:odd}}

Octupole correlations are expected to be 
relevant to the low-lying states of odd-mass 
nuclei, as well as to the even-even ones. 
To treat odd-mass nuclear systems, 
coupling between 
a single-particle (or unpaired nucleon) 
degree of freedom and collective 
even-even boson space is considered 
within the interacting 
boson-fermion model (IBFM) \cite{IBFM,iachello1979}. 
The IBFM Hamiltonian consists of the 
IBM Hamiltonian $\hat H_\text{B}$, 
single-nucleon 
Hamiltonian $\hat H_\text{F}$, and 
a term $\hat V_\text{BF}$ 
representing the boson-fermion coupling: 
\begin{eqnarray}
\label{eq:fham}
 \hat H_\text{IBFM} = \hat H_\text{B} + \hat H_\text{F} + \hat V_\text{BF} \; .
\end{eqnarray}
Here, 
$\hat H_\text{B}$ was defined in Eq.~(\ref{eq:bham}), 
and $\hat H_\text{F} = \sum_j \epsilon_j 
a_j^\dagger \cdot \tilde a_{j}$, 
with $a_j^{(\dagger)}$ and $\epsilon_j$ 
being the annihilation (creation) operator 
of a nucleon in the orbital $j$ and 
the single-particle energy, respectively. 
The interaction $\hat V_\text{BF}$ 
consists of the monopole, dynamical quadrupole 
and octupole, and exchange terms 
\cite{nomura2018oct}. 
By using the generalized seniority scheme 
\cite{scholten1985,nomura2018oct}, 
coefficients of the boson-fermion interaction 
terms are shown to be given as functions of 
particle occupation numbers ($v^2_j$).

As an illustrative example, 
let us consider the nucleus $^{145}$Ba, 
a system of a single neutron coupled 
to the even-even nucleus $^{144}$Ba. 
Experimentally, 
a static octupole deformation has been 
identified in $^{144}$Ba \cite{bucher2016}, 
while the octupole correlations have been 
suggested to be present in the odd-mass neighbor 
$^{145}$Ba \cite{rzacaurban2012}. 
For the single-particle space of the 
odd neutron we include the entire 
$N$ = 82-126 neutron major oscillator shell 
consisting of the 
$3p_{1/2}$, $3p_{3/2}$, $2f_{5/2}$, $2f_{7/2}$, 
$1h_{9/2}$, and $1i_{13/2}$ orbits.

The procedure to fix the IBFM Hamiltonian 
in Eq.~(\ref{eq:fham}) goes as follows 
\cite{nomura2016odd,nomura2018oct}. 
First, the $sdf$-IBM Hamiltonian $\hat H_\text{B}$ 
is determined by the $\beta_2$-$\beta_3$ 
energy-surface mapping as described 
in Sec.~\ref{sec:mapping}. 
Second, the spherical single-particle 
energies $\epsilon_j$ 
and occupation probabilities $v^2_j$, which enter 
$\hat H_\text{F}$ and $\hat V_\text{BF}$, 
respectively, are provided 
by the SCMF calculation constrained to zero 
quadrupole and octupole deformations. 
Finally, only the strength parameters for 
the boson-fermion interactions are treated 
as free parameters, and are fitted to reproduce 
the empirical low-energy spectra of the odd-mass 
nucleus.

\begin{figure}[th]
\centerline{\includegraphics[width=.6\linewidth]{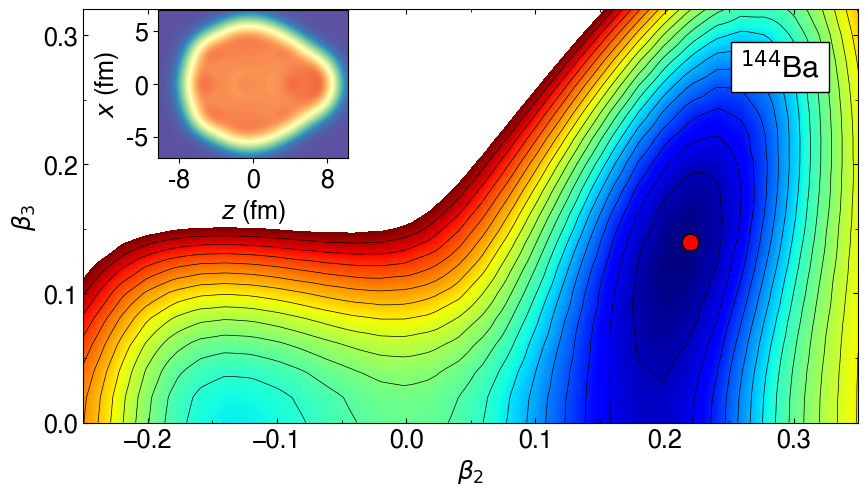}}
\caption{Same as Fig.~\ref{fig:pes-th}, but for $^{144}$Ba. }
\label{fig:pes-ba}
\end{figure}

Figure~\ref{fig:pes-ba} shows
SCMF axially-symmetric 
$\beta_2$-$\beta_3$ energy 
surface for the even-even core $^{144}$Ba, 
obtained from the RHB method that employs 
the functional DD-PC1 
and the separable pairing force. 
One clearly sees an octupole deformed minimum 
at ($\beta_2$, $\beta_3$) $\approx$ (0.22, 0.14), 
suggesting a pronounced octupole collectivity. 
The intrinsic nucleon density projected 
on the $x$-$z$ plane, 
also shown in Fig.~\ref{fig:pes-ba}, 
indeed resembles a reflection-asymmetric 
nuclear shape. 

\begin{figure}[th]
\centerline{\includegraphics[width=.8\linewidth]{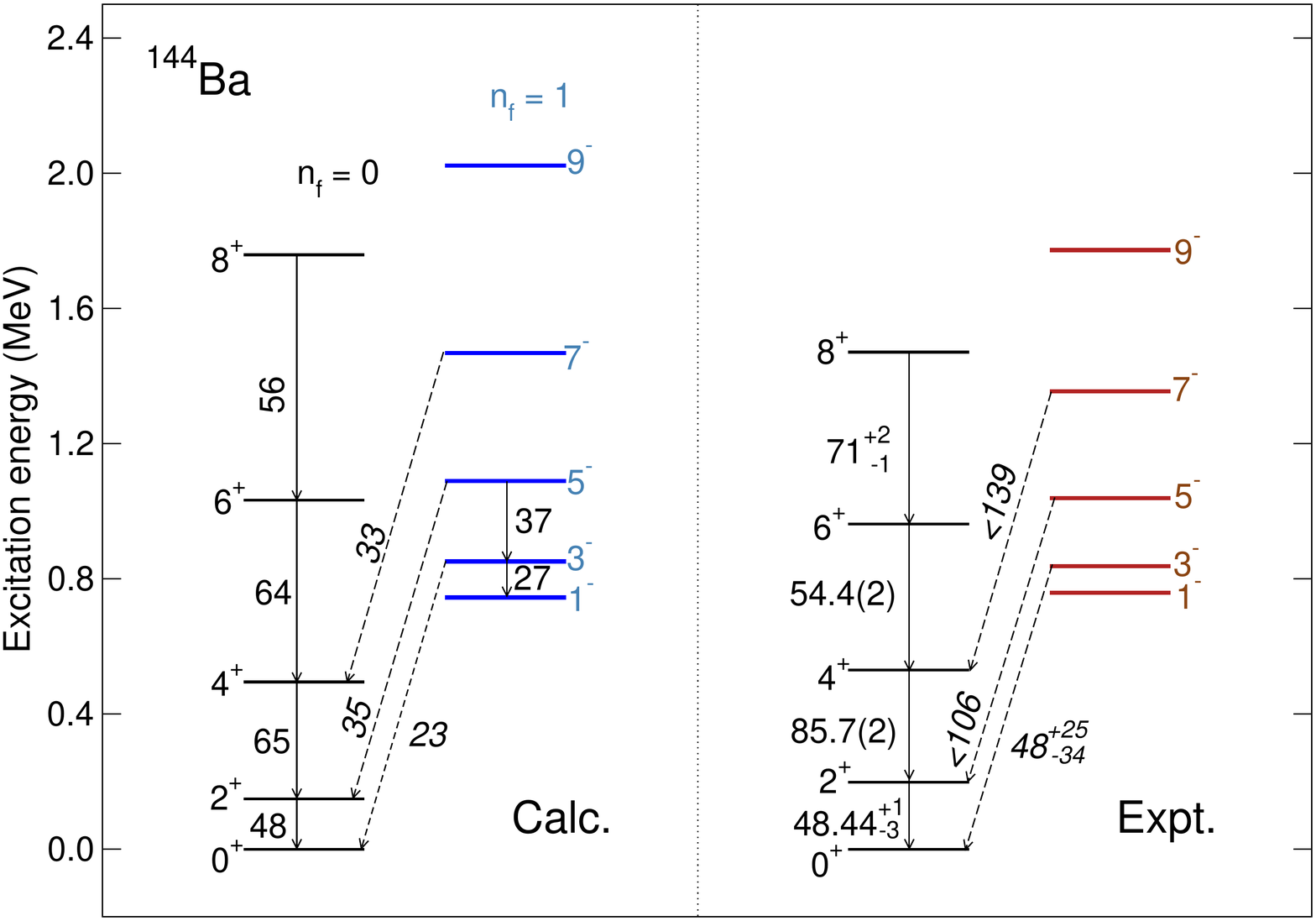}}
\caption{Calculated and experimental \cite{bucher2016} 
low-energy 
excitation spectra for positive- and negative-parity 
yrast states of $^{144}$Ba. Numbers along solid and 
dotted lines are $B(E2)$ and $B(E3)$ values 
in W.u., respectively. The calculated positive- and 
negative-parity states are made of zero ($n_f$ = 0) 
and one ($n_f$ = 1) $f$ boson configurations, 
respectively. The DD-PC1 EDF is employed 
for the microscopic input to the $sdf$-IBM.}
\label{fig:ba144}
\end{figure}

In Fig.~\ref{fig:ba144}, 
low-energy excitation spectra for the 
positive- and negative-parity yrast states 
of $^{144}$Ba resulting from the $sdf$-IBM 
are compared with the experimental data 
\cite{bucher2016}. 
For states belonging to the negative-parity 
yrast band, the matrix element 
of the $f$-boson number operator 
is calculated to be 
$\braket{\hat n_f}$ $\approx$ 1, 
that is, the band is mainly 
constructed by one-$f$ boson ($n_f$ = 1) 
components coupled with the $sd$ boson space. 
The positive-parity 
band is, however, solely made of the $sd$ 
bosons, as $\braket{n_f}$ $\approx$ 0. 
The calculated $B(E3; 3^-_1 \to 0^+_1)$ 
transition probability of 23 W.u. 
is comparable to the measured value, 
48$^{+25}_{-34}$ W.u.

\begin{figure}[th]
\centerline{\includegraphics[width=\linewidth]{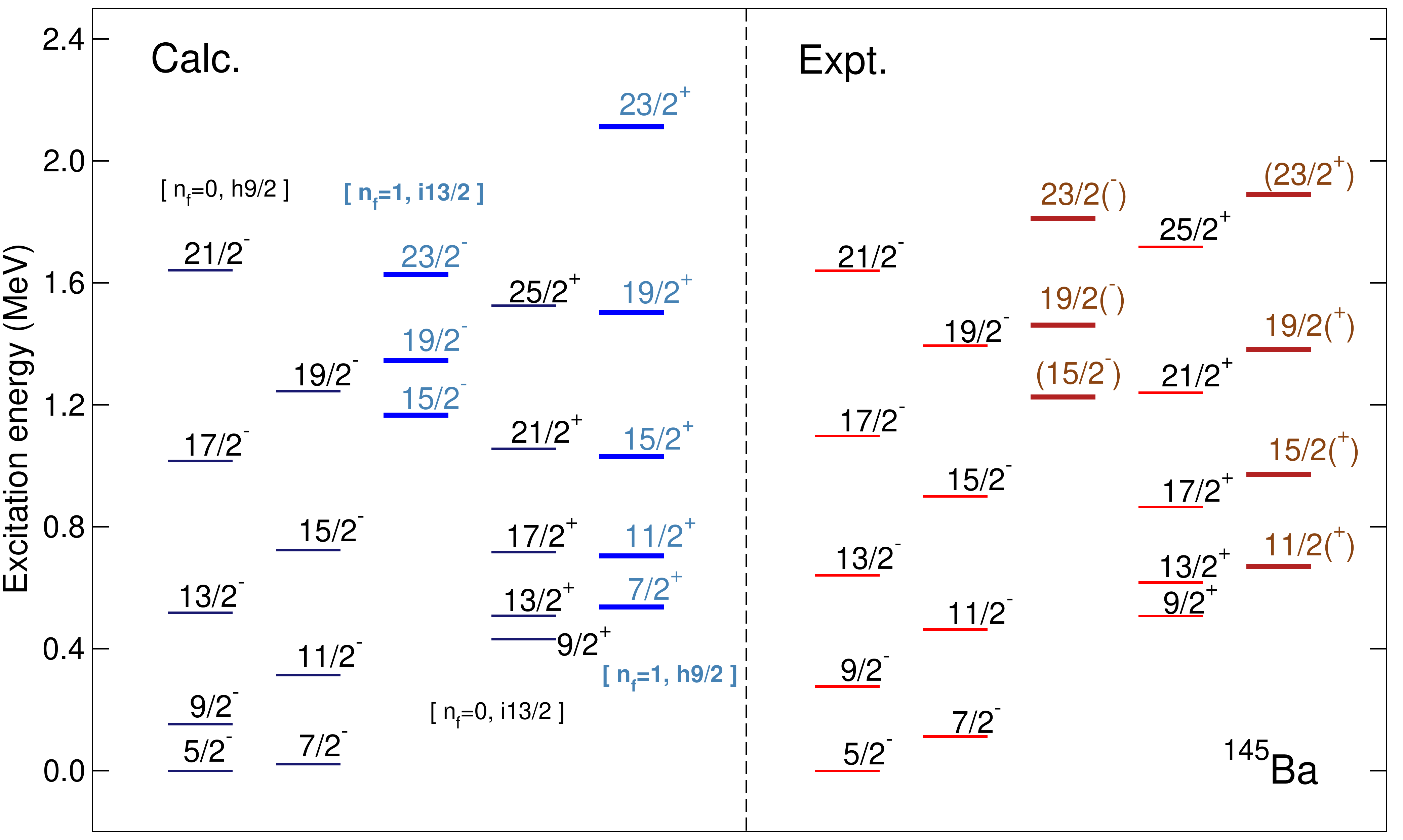}}
\caption{Calculated and experimental 
\cite{rzacaurban2012} energy spectra 
for the low-lying 
negative- and positive-parity states for 
the odd-$A$ nucleus $^{145}$Ba. Those 
bands that are suggested to be of octupole nature 
are highlighted in thick lines. The calculations 
are performed within the $sdf$-IBFM 
that is based on the universal energy 
functional DD-PC1.}
\label{fig:ba145}
\end{figure}

The low-energy excitation spectra 
for the neighboring odd-mass nucleus $^{145}$Ba 
computed by the $sdf$-IBFM are given 
in Fig.~\ref{fig:ba145}. 
The $sdf$-IBFM wave functions 
for the lowest two  
${5/2}^-$ and ${7/2}^-$ negative-parity band 
are mainly composed of the configuration 
of the odd neutron in the $1h_{9/2}$ orbit 
coupled to the $sd$ boson space, with 
the matrix element $\braket{n_f}$ $\approx$ 0. 
The lowest positive-parity band with the 
bandhead ${9/2}^+_1$ is mostly made of 
the odd neutron in the $1i_{13/2}$ 
unique-parity orbit coupled to the 
$sd$ boson space. 
Experimentally \cite{rzacaurban2012} 
the observed $\Delta I$ = 2 negative-parity 
band built on the ${15/2}^-$ state at 1226 keV, 
and positive-parity band built on the ${11/2}^+$ 
state at 670 keV 
are interpreted as octupole bands. 
Corresponding theoretical bands 
are the ones based on the 
${15/2}^-_4$ and ${7/2}^+_1$ states. 
These are, respectively, determined by 
the configurations of the odd neutron in the 
$1i_{13/2}$ and $1h_{9/2}$ orbits 
coupled to the boson space 
containing one $f$ boson 
($\braket{n_f}$ $\approx$ 1). 
The predicted octupole bands 
have rather large $E3$ transitions to 
non-octupole ($\braket{n_f}$ $\approx$ 0) states: 
$B(E3; {15/2}^-_4 \to {9/2}^+_1)$ = 25, 
$B(E3; {19/2}^-_2 \to {13/2}^+_1)$ = 31,  
$B(E3; {11/2}^+_1 \to {5/2}^-_1)$ = 4.7, and 
$B(E3; {15/2}^+_1 \to {9/2}^-_1)$ = 15 
(all in W.u.)

\subsection{Shape coexistence and low-lying $0^+$ states}

Shape coexistence
is an intriguing feature of nuclear 
structure, in which different 
intrinsic structures 
appear near the ground state of a single nucleus 
\cite{wood1992,heyde2011}. 
Manifestation of the shape coexistence 
is the existence of low-lying $0^+$ excited 
states close in energy to the ground state. 
The extra low-lying $0^+$ states are 
interpreted as intruder states 
arising from the cross-shell 
multiparticle-multihole excitations. 
With the particle-hole excitations 
the residual neutron-proton correlations 
are increased 
to such a degree as to lower the $0^+$ energies 
\cite{federman1977,heyde1987}. 
The different intruder $0^+$ states 
and subsequent bands 
can also be associated with 
different intrinsic deformations 
in the mean-field potential energy surface 
\cite{bengtsson1987,nazarewicz1993}.

To incorporate the intruder states 
in the IBM, 
one can adopt the prescription proposed  
by Duval and Barrett \cite{duval1981}, 
in which 
the usual IBM Hilbert space is extended to 
be a direct sum of the shell-model-like 
0p-0h normal, and 2p-2h, 4p-4h, $\ldots$ 
intruder configuration spaces. 
Several independent IBM Hamiltonians, 
corresponding to different configurations,  
are introduced and are allowed to be mixed. 
Assuming that the particle and hole states 
are not distinguished, these Hamiltonians 
differ in boson number by two. 
The total IBM Hamiltonian 
that is to carry out the configuration 
mixing of the 2$k$p-2$k$h ($k$ = 0, 1, 2, $\ldots$) 
states is then given by
\begin{eqnarray}
\label{eq:cmham}
 \hat H = \hat P_0 \hat H_{0} \hat P_0
+ \sum_{k=1} \hat P_k ( \hat H_{k} + \Delta_{k} ) \hat P_k 
+ \sum_{k=0} \hat P_{k+1} \hat V_{k,k+1} \hat P_k 
+ (H.c.)
\; .
\end{eqnarray}
$\hat H_k$ is the unperturbed Hamiltonian 
for the 2$k$p-2$k$h configuration, 
the operator $\hat P_k$ projects 
states onto the 2$k$p-2$k$h space, 
and the constant $\Delta_k$ 
($k$ $\geq$ 1) represents 
the energy required to promote 
$2k$ nucleons (or $k$ bosons) from one 
major shell to another. 
The interaction $\hat V_{k,k+1}$ admixes 
the 2$k$p-2$k$h and 2$(k+1)$p-2$(k+1)$h 
configurations, hence does not conserve 
boson number.

Since the configuration mixing IBM framework  
significantly alleviates the computational 
burden often accompanying massive computations  
as in the large-scale shell model, 
it has been extensively employed 
as an alternative approach to 
shape coexistence. 
A number of phenomenological studies 
using the configuration-mixing IBM 
have been carried out, e.g., 
for the neutron-deficient Hg 
\cite{garciaramos2014}, neutron-rich Zr 
\cite{gavrielov2019}, 
and even-even Cd \cite{leviatan2018}. 
On the other hand, 
since several independent IBM Hamiltonians 
and mixing terms are considered, 
this framework naturally 
involves a large number of 
adjustable parameters.

The method of deriving the 
configuration-mixing 
IBM Hamiltonian from the SCMF calculation  
has been developed \cite{nomura2012sc} 
for the description of the quadrupole 
collective states in the Pb region. 
More recently, this method was extended to include 
the $f$ boson degrees of freedom so as to 
deal with the shape coexistence that involves 
both quadrupole and octupole states \cite{nomura2022octcm}. 
The following discussion concerns 
this new development, taking the nucleus 
$^{76}$Kr as a representative case. 
Note that there are a few instances in which 
the configuration mixing $sdf$ (or $spdf$) 
IBM has been used in a purely phenomenological 
way for describing structure of 
even-even Cd isotopes 
\cite{Garrett1999-112Cd,cd114-2003}.

\begin{figure}[th]
\centerline{\includegraphics[width=\linewidth]{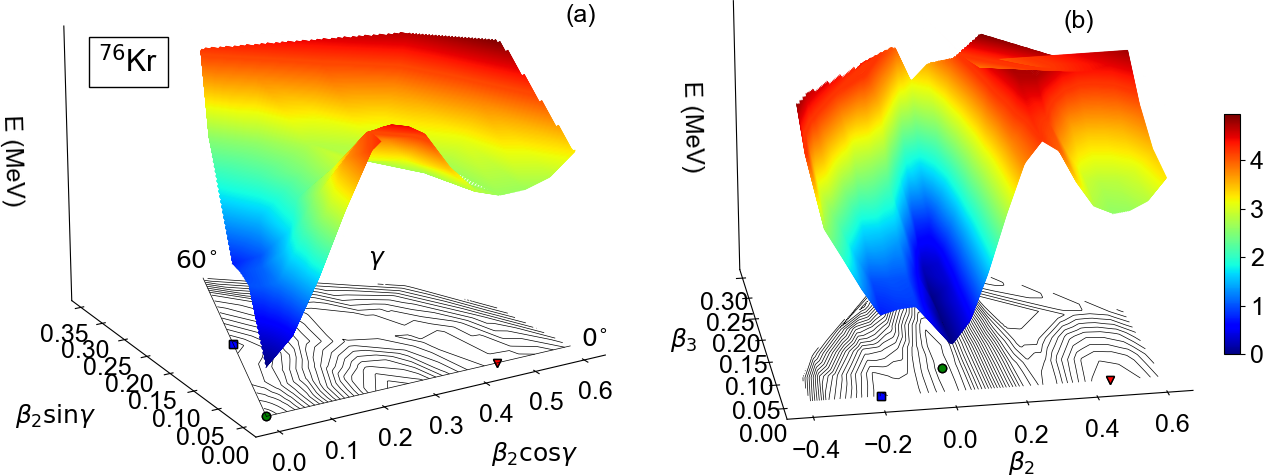}}
\caption{Potential energy surfaces for $^{76}$Kr 
in terms of the (a) triaxial quadrupole ($\beta_2$,$\gamma$) 
and (b) axially-symmetric quadrupole $\beta_2$ 
and octupole $\beta_3$ deformations, 
calculated within the constrained RHB 
method based on the DD-PC1 functional 
and separable pairing force. 
Two dimensional 
contour plots are also shown, with the 
energy difference between neighboring 
contours being 0.2 MeV. 
The symbols 
circle, square, and triangle represent 
the spherical global minimum, and oblate 
and prolate local minima, respectively.}
\label{fig:pes-kr76}
\end{figure}

To begin with, 
two sets of SCMF calculations are 
performed, with  
constraints on the triaxial quadrupole 
($\beta_2$, $\gamma$) deformations 
with $\beta_3$ = 0, and on the 
axially symmetric quadrupole-octupole 
($\beta_2$, $\beta_3$) deformations. 
The angle variable $\gamma$ 
corresponds to the triaxial deformation, 
and is defined within the range 
0$^\circ$ $\leqslant$ $\gamma$ $\leqslant$ 60$^\circ$. 
Figure~\ref{fig:pes-kr76} shows 
the potential energy 
surfaces within the ($\beta_2$, $\gamma$) 
and ($\beta_2$, $\beta_3$) planes, 
computed by the RHB method 
using the functional DD-PC1 and the 
separable paring force. 
In the $\beta_2$-$\gamma$ energy surface, 
a spherical global minimum is obtained and, 
in addition, an weakly oblate deformed and a strongly 
prolate deformed local minima can be seen at 
($\beta_2$, $\gamma$) $\approx$ (0.2, 60$^\circ$) 
and (0.45, 0$^{\circ}$), respectively. 
The spherical ground state reflects the 
effect of the neutron $N=40$ subshell closure. 
The $\beta_2$-$\beta_3$ energy surface 
has the global minimum at  
($\beta_2$, $\beta_3$) $\approx$ (0.0, 0.05) 
with non-zero octupole deformation, 
while the potential is soft along 
the $\beta_3$ deformation.

Since the energy surface 
exhibits three minima, 
the IBM space consists of 
the three configurations corresponding 
to the 0p-0h, 2p-2h, and 4p-4h states 
that comprise $n$, $n+2$, and $n+4$ bosons. 
The normal 0p-0h space here corresponds 
to the $N/Z$ = 28-50 major shell, 
hence the boson number $n$ = 9. 
In addition, 
the unperturbed 0p-0h Hamiltonian 
is associated with the spherical 
global minimum on the energy surfaces, 
while the 2p-2h and 4p-4h Hamiltonians 
are assigned to the oblate 
and prolate local minima, respectively. 
The full configuration mixing $sdf$ IBM 
Hamiltonian is then determined by the 
procedure described in 
Ref.~\refcite{nomura2022octcm}. 
The corresponding Hamiltonian is 
diagonalized in the boson space
\begin{eqnarray}
 [(sdf)^n] \oplus [(sdf)^{n+2}] \oplus [(sdf)^{n+4}] \; .
\end{eqnarray}

\begin{figure}[th]
\centerline{\includegraphics[width=\linewidth]{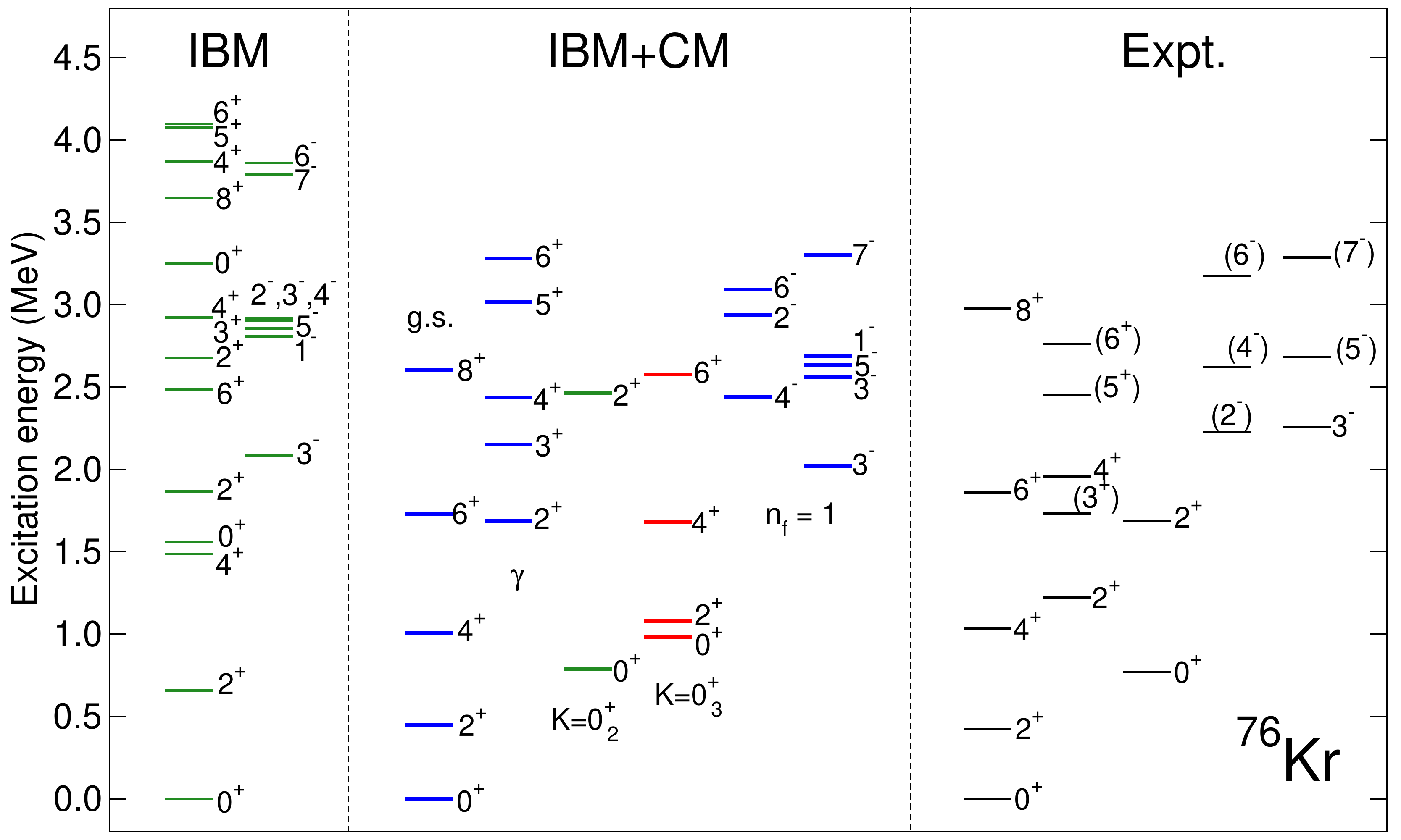}}
\caption{Excitation spectra for 
positive- and negative-parity states for $^{76}$Kr 
calculated by using a single configuration (``IBM'')
and taking into account the configuration mixing (``IBM+CM''). 
The IBM calculations are performed 
with the microscopic basis on the 
universal functional DD-PC1. 
Experimental data \cite{clement2007,data} 
are shown on the right.}
\label{fig:kr76}
\end{figure}

In the middle part 
of Fig.~\ref{fig:kr76} the low-energy 
excitation spectra for $^{76}$Kr, 
resulting from 
the configuration mixing $sdf$-IBM, 
are compared with the 
experimental data \cite{clement2007,data},  
and with the usual $sdf$-IBM result 
obtained by using the single configuration 
associated with the spherical ground 
state. 
In both sets of the $sdf$-IBM, 
all the negative-parity states are based on the 
one-$f$ boson configuration with the expectation 
value $\braket{\hat n_f}$ $\approx$ 1, 
while $\braket{\hat n_f}$ $\approx$ 0 
for the positive parity states.

A notable effect of taking into account the 
configuration mixing is the lowering in energy 
of the excited $0^+$ states. 
In the usual $sdf$-IBM 
with the single configuration, 
the $0^+_2$ and $0^+_3$ excitation 
energies are 
calculated to be $E_x$ $\approx$ 1.5 MeV and 3.25 MeV, 
respectively, whereas 
they are both $E_x$ $<$ 1 MeV 
in the configuration mixing 
IBM. 
The energy levels for the positive-parity 
yrast states with spin $I>0$ 
and also for the negative-parity states 
are generally lowered 
as a result of the mixing. 
A near degeneracy 
of the $1^-_1$, $2^-_1$, $3^-_2$, $4^-_1$, and $5^-_1$ levels, 
obtained in the single-configuration 
calculation, can be interpreted as 
a quintet arising from the quadrupole-octupole 
phonon coupling $2^+ \otimes 3^-$. 
The degeneracy is, however, removed in the 
configuration mixing calculation as a 
consequence of the level repulsion.

Regarding the transition properties, 
the $B(E3; 3^-_1 \to 0^+_1)$ 
rates are calculated to be 
$\approx$ 30 W.u. in this mass 
region \cite{nomura2022octcm}
An effect of the configuration mixing 
is, for instance, increase of the ratio 
$B(E3; 3^-_1 \to 2^+_1)/B(E3; 3^-_1 \to 0^+_1)$ 
by an order of magnitude. 
However, experimental data 
for the $E3$ and $E1$ 
transitions in this region are scarce, 
and a more thorough assessment of the 
model remains to be done.

\begin{figure}[th]
\centerline{\includegraphics[width=\linewidth]{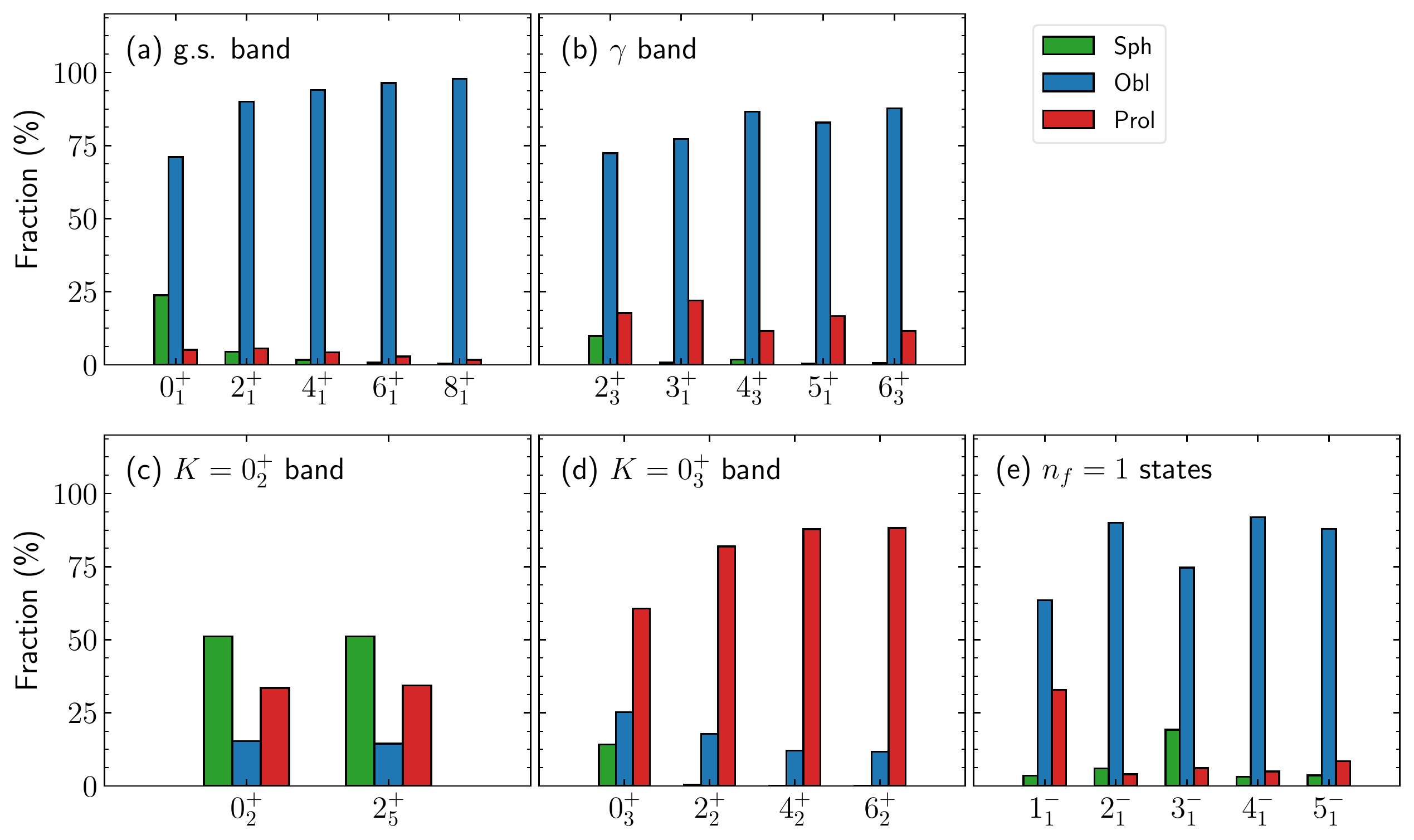}}
\caption{Compositions of the spherical, 
oblate, and prolate configurations in the 
wave functions of the positive-parity 
states belonging to the 
(a) ground-state, (b) $\gamma$-, (c) $K^\pi=0^+_2$, 
and (d) $K^\pi=0^+_3$ bands, and of the 
(e) low-spin negative-parity 
(one $f$ boson, $n_f$ = 1) states for $^{76}$Kr.}
\label{fig:wf-kr76}
\end{figure}

To analyze in more detail the nature 
of the proposed low-spin bands, 
Fig.~\ref{fig:wf-kr76} shows compositions 
of the spherical (0p-0h), oblate (2p-2h), 
and prolate (4p-4h) configurations in 
the wave functions of the members of these bands. 
Even though the spherical ground state is 
predicted at the SCMF level, 
at the IBM level 
the oblate 2p-2h configuration 
constitutes most of the wave functions for 
the ground-state and $\gamma$ bands, and all 
the negative-parity states shown in 
Fig.~\ref{fig:kr76}. 
The $K^\pi$ = $0^+_2$ band with the bandhead 
$0^+_2$ state is, in turn, spherical in nature. 
However, substantial amounts of 
the oblate and prolate components 
are admixed into the members of the proposed 
$K^\pi$ = $0^+_2$ band. 
The $K^\pi$ = $0^+_3$ rotational-like band 
based upon the $0^+_3$ state is dominated 
by the strongly deformed prolate 
configuration.

\subsection{Double octupole phonon states}

A number of low-lying excited 
$0^+$ states are also observed in 
those nuclei in rare-earth region 
\cite{aprahamian2018}. 
Microscopic origin of 
these $0^+$ states is under active debate, 
and has been attributed, e.g., to the pairing 
vibrations \cite{xiang2020,nomura2020pv}, 
intruder excitations \cite{vanisacker1982}, and 
coupling of the double octupole phonons 
\cite{zamfir2002}. 
To explain 
the occurrence of these low-energy $0^+$ states 
the $sdf$- or $spdf$-IBM has often been used. 
In particular, 
calculation within 
the $sdf$-IBM that is based on the Gogny EDF 
has shown \cite{nomura2015} 
that the $0^+_2$ excited states 
in many of the Sm and Gd nuclei 
are formed mainly by the coupling 
of two $f$ bosons. 
The double octupole phonon nature 
of the $0^+_2$ state has been suggested 
in light actinides as well within 
the $sdf$-IBM calculation 
based on the relativistic EDF 
\cite{nomura2014}.

\subsection{Coupling with $p$ boson}

A straightforward extension of the $sdf$-boson model 
would be to include the dipole $p$ boson. 
Inclusion of this boson degree of freedom 
would indeed improve the description of 
the negative-parity levels and their 
$E1$ transition properties 
\cite{sugita1996}, and has also been 
considered in the calculations on 
the $\alpha$-clustering phenomena 
\cite{iachello1982,daley1983,spieker2015,bijker2017}. 
At the microscopic level, 
it was shown \cite{otsuka1986},  
by using the Nilsson model, 
that large amount of the dipole 
pair, i.e., $p$ boson, component is contained in 
the wave functions of negative-parity states, 
and that the dipole boson 
is as essential a degree of freedom 
as the octupole one. 
The physical meaning of the $p$ 
boson has been 
attributed to the center of mass 
motion or giant dipole resonance 
\cite{otsuka1986,otsuka1988}, but 
is not as entirely clear as that of 
the $f$ boson, 
since the dipole mode is more or less of 
single-particle nature.

\section{Conclusions}

We have reviewed recent theoretical 
investigations of the 
octupole shapes and collective excitations 
in nuclei within the $sdf$-IBM 
that is formulated microscopically 
using the nuclear EDF framework. 
The illustrative application to Th isotopes 
has shown the occurrence of the 
stable octupole state around $N$ = 134 
as well as the shape phase transition 
that involves both quadrupole and 
octupole modes. 
A global spectroscopic study  
has confirmed the enhancement 
of the octupole 
collectivity in a wide mass region. 
The coupling between the octupole and 
additional degrees of freedom has been 
shown to play a role 
in the descriptions of a variety of 
related nuclear structure 
phenomena, including the shape coexistence.

The present model descriptions 
are based on the assumptions that the 
nuclear EDFs, upon which the IBM 
Hamiltonian is built, provide for all nuclei 
the correct mean-field properties  
(potential energy surfaces, 
single-particle energies, etc.), 
and that the EDF-to-IBM mapping procedure 
is valid. 
In this respect, 
for a more refined description of 
the octupole-related properties, 
further improvements of the method 
would be made, e.g., through (i) the assessment 
of the capability 
of a given EDF of predicting complex nuclear 
shape phenomena as well as spectroscopy, 
(ii) the inclusions of additional terms in the 
boson Hamiltonian and/or degrees of 
freedom, such as the $p$ boson, 
in the boson space, 
and (iii) the microscopic derivation 
of the boson effective charges for 
the $E1$ and $E3$ transition operators. 

Furthermore, 
the theoretical method discussed 
throughout this paper has a potential to 
explore problems in interdisciplinary 
fields. For instance, 
the semi-microscopic $sdf$-IBFM, described in 
Sec.~\ref{sec:odd}, can be used to study 
the proposed parity doublet and EDM for  
the odd-mass Ra and Th isotopes, 
which serve as a testing ground 
for the CP symmetry violation 
\cite{dobaczewski2005}.

\section*{Acknowledgements}

The author thanks D. Vretenar, T. Nik\v{s}i\'c, 
L. M. Robledo, and R. Rodr\'iguez-Guzm\'an for 
their contributions to the works discussed 
in this paper. The author has been 
supported by the Tenure Track Pilot Programme of 
the Croatian Science Foundation and 
the \'Ecole Polytechnique 
F\'ed\'erale de Lausanne, and 
the Project TTP-2018-07-3554 Exotic 
Nuclear Structure and Dynamics, 
with funds of the Croatian-Swiss 
Research Programme.

\bibliographystyle{ws-ijmpe}
\bibliography{sample}

\begin{thebibliography}{10}

\bibitem{butler1996}
P.~A. Butler and W.~Nazarewicz, {\em Rev. Mod. Phys.} {\bf 68}  (1996) 349.

\bibitem{butler2016}
P.~A. Butler, {\em J. Phys. G: Nucl. Part. Phys.} {\bf 43}  (2016)   073002.

\bibitem{gaffney2013}
L.~P. {Gaffney {\it{et al.}}}, {\em Nature (London)} {\bf 497}  (2013) 199.

\bibitem{chishti2020}
M.~M.~R. {Chishti \it{et al.}}, {\em Nat. Phys.} {\bf 16}  (2020) 853.

\bibitem{butler2020a}
P.~A. {Butler \it{et al.}}, {\em Phys. Rev. Lett.} {\bf 124}  (2020)   042503.

\bibitem{bucher2016}
B.~{Bucher \it{et al.}}, {\em Phys. Rev. Lett.} {\bf 116}  (2016)   112503.

\bibitem{bucher2017}
B.~{Bucher \it{et al.}}, {\em Phys. Rev. Lett.} {\bf 118}  (2017)   152504.

\bibitem{engel2013}
J.~Engel, M.~J. Ramsey-Musolf and U.~{van Kolck}, {\em Prog. Part. Nucl. Phys.}
  {\bf 71}  (2013) 21 .

\bibitem{IBM}
F.~Iachello and A.~Arima, {\em The interacting boson model} (Cambridge Univ.
  Press, Cambridge, UK, 1987).

\bibitem{cejnar2010}
P.~Cejnar, J.~Jolie and R.~F. Casten, {\em Rev. Mod. Phys.} {\bf 82}  (2010)
  2155.

\bibitem{wood1992}
J.~L. Wood, K.~Heyde, W.~Nazarewicz, M.~Huyse and P.~van Duppen, {\em Phys.
  Rep.} {\bf 215}  (1992) 101 .

\bibitem{heyde2011}
K.~Heyde and J.~L. Wood, {\em Rev. Mod. Phys.} {\bf 83}  (2011) 1467.

\bibitem{arima1975}
A.~Arima and F.~Iachello, {\em Phys. Rev. Lett.} {\bf 35}  (1975) 1069.

\bibitem{OAIT}
T.~Otsuka, A.~Arima, F.~Iachello and I.~Talmi, {\em Phys. Lett. B} {\bf 76}
  (1978) 139 .

\bibitem{OAI}
T.~Otsuka, A.~Arima and F.~Iachello, {\em Nucl. Phys. A} {\bf 309}  (1978)  ~1.

\bibitem{arima1978su3}
A.~Arima and F.~Iachello, {\em Ann. Phys. (NY)} {\bf 111}  (1978) 201 .

\bibitem{engel1985}
J.~Engel and F.~Iachello, {\em Phys. Rev. Lett.} {\bf 54}  (1985) 1126.

\bibitem{engel1987}
J.~Engel and F.~Iachello, {\em Nucl. Phys. A} {\bf 472}  (1987) 61 .

\bibitem{barfield1988}
A.~F. {Barfield \it{et al.}}, {\em Ann. Phys.} {\bf 182}  (1988) 344 .

\bibitem{zamfir2001}
N.~V. Zamfir and D.~Kusnezov, {\em Phys. Rev. C} {\bf 63}  (2001)   054306.

\bibitem{cottle1996}
P.~D. Cottle and N.~V. Zamfir, {\em Phys. Rev. C} {\bf 54}  (1996) 176.

\bibitem{kusnezov1988}
D.~Kusnezov and F.~Iachello, {\em Phys. Lett. B} {\bf 209}  (1988) 420 .

\bibitem{kusnezov1989}
D.~Kusnezov, {\em J. Phys. A: Math. Gen.} {\bf 22}  (1989) 4271.

\bibitem{kusnezov1990}
D.~Kusnezov, {\em J. Phys. A: Math. Gen.} {\bf 23}  (1990) 5673.

\bibitem{mizusaki1996}
T.~Mizusaki and T.~Otsuka, {\em Prog. Theor. Phys. Suppl.} {\bf 125}  (1996)
  97.

\bibitem{nomura2008}
K.~Nomura, N.~Shimizu and T.~Otsuka, {\em Phys. Rev. Lett.} {\bf 101}  (2008)
  142501.

\bibitem{nomura2010}
K.~Nomura, N.~Shimizu and T.~Otsuka, {\em Phys. Rev. C} {\bf 81}  (2010)
  044307.

\bibitem{nomura2011rot}
K.~Nomura, T.~Otsuka, N.~Shimizu and L.~Guo, {\em Phys. Rev. C} {\bf 83}
  (2011)   041302.

\bibitem{nomura2012tri}
K.~Nomura, N.~Shimizu, D.~Vretenar, T.~Nik\ifmmode \check{s}\else
  \v{s}\fi{}i\ifmmode~\acute{c}\else \'{c}\fi{} and T.~Otsuka, {\em Phys. Rev.
  Lett.} {\bf 108}  (2012)   132501.

\bibitem{nomura2012sc}
K.~Nomura, R.~Rodr\'{\i}guez-Guzm\'an, L.~M. Robledo and N.~Shimizu, {\em Phys.
  Rev. C} {\bf 86}  (2012)   034322.

\bibitem{nomura2013oct}
K.~Nomura, D.~Vretenar and B.-N. Lu, {\em Phys. Rev. C} {\bf 88}  (2013)
  021303.

\bibitem{nomura2014}
K.~Nomura, D.~Vretenar, T.~Nik\ifmmode \check{s}\else
  \v{s}\fi{}i\ifmmode~\acute{c}\else \'{c}\fi{} and B.-N. Lu, {\em Phys. Rev.
  C} {\bf 89}  (2014)   024312.

\bibitem{nomura2015}
K.~Nomura, R.~Rodr\'{\i}guez-Guzm\'an and L.~M. Robledo, {\em Phys. Rev. C}
  {\bf 92}  (2015)   014312.

\bibitem{nomura2016odd}
K.~Nomura, T.~Nik\ifmmode \check{s}\else \v{s}\fi{}i\ifmmode~\acute{c}\else
  \'{c}\fi{} and D.~Vretenar, {\em Phys. Rev. C} {\bf 93}  (2016)   054305.

\bibitem{nomura2020beta-2}
K.~Nomura, R.~Rodr\'{\i}guez-Guzm\'an and L.~M. Robledo, {\em Phys. Rev. C}
  {\bf 101}  (2020)   044318.

\bibitem{nomura2020beta-1}
K.~Nomura, R.~Rodr\'{\i}guez-Guzm\'an and L.~M. Robledo, {\em Phys. Rev. C}
  {\bf 101}  (2020)   024311.

\bibitem{bender2003}
M.~Bender, P.-H. Heenen and P.-G. Reinhard, {\em Rev. Mod. Phys.} {\bf 75}
  (2003)   121.

\bibitem{vretenar2005}
D.~Vretenar, A.~V. Afanasjev, G.~A. Lalazissis and P.~Ring, {\em Phys. Rep.}
  {\bf 409}  (2005) 101 .

\bibitem{niksic2011}
T.~Nik\ifmmode \check{s}\else \v{s}\fi{}i\ifmmode~\acute{c}\else \'{c}\fi{},
  D.~Vretenar and P.~Ring, {\em Prog. Part. Nucl. Phys.} {\bf 66}  (2011)
  519.

\bibitem{robledo2019}
L.~M. Robledo, T.~R. Rodríguez and R.~R. Rodr{\'{i}}guez-Guzm{\'{a}}n, {\em J.
  Phys. G: Nucl. Part. Phys.} {\bf 46}  (2019)   013001.

\bibitem{schunck2019}
N.~Schunck (ed.), {\em Energy Density Functional Methods for Atomic Nuclei}
  (IOP Publishing, 2019).

\bibitem{RS}
P.~Ring and P.~Schuck, {\em The Nuclear Many-Body Problem} (Springer, Berlin,
  1980).

\bibitem{DDPC1}
T.~Nik\ifmmode \check{s}\else \v{s}\fi{}i\ifmmode~\acute{c}\else \'{c}\fi{},
  D.~Vretenar and P.~Ring, {\em Phys. Rev. C} {\bf 78}  (2008)   034318.

\bibitem{tian2009}
Y.~Tian, Z.~Y. Ma and P.~Ring, {\em Phys. Lett. B} {\bf 676}  (2009) 44 .

\bibitem{li2016}
Z.~P. Li, T.~Nik{\v{s}}i{\'{c}} and D.~Vretenar, {\em J. Phys. G: Nucl. Part.
  Phys.} {\bf 43}  (2016)   024005.

\bibitem{ginocchio1980}
J.~N. Ginocchio and M.~W. Kirson, {\em Nucl. Phys. A} {\bf 350}  (1980)  ~31.

\bibitem{dieperink1980}
A.~E.~L. Dieperink, O.~Scholten and F.~Iachello, {\em Phys. Rev. Lett.} {\bf
  44}  (1980) 1747.

\bibitem{bohr1980}
A.~Bohr and B.~R. Mottelson, {\em Phys. Scr.} {\bf 22}  (1980) 468.

\bibitem{nomura2021qoch}
K.~Nomura, L.~Lotina, T.~Nik\ifmmode \check{s}\else
  \v{s}\fi{}i\ifmmode~\acute{c}\else \'{c}\fi{} and D.~Vretenar, {\em Phys.
  Rev. C} {\bf 103}  (2021)   054301.

\bibitem{nomura2020oct}
K.~{Nomura \it{et al.}}, {\em Phys. Rev. C} {\bf 102}  (2020)   064326.

\bibitem{li2013}
Z.~P. {Li \it{et al.}}, {\em Phys. Lett. B} {\bf 726}  (2013) 866 .

\bibitem{xia2017}
S.~Y. {Xia \it{et al.}}, {\em Phys. Rev. C} {\bf 96}  (2017)   054303.

\bibitem{data}
{Brookhaven National Nuclear Data Center}. {http://www.nndc.bnl.gov}.

\bibitem{nomura2021oct-u}
K.~{Nomura \it{et al.}}, {\em Phys. Rev. C} {\bf 103}  (2021)   044311.

\bibitem{nomura2021oct-ba}
K.~{Nomura \it{et al.}}, {\em Phys. Rev. C} {\bf 104}  (2021)   044324.

\bibitem{nomura2021oct-zn}
K.~{Nomura \it{et al.}}, {\em Phys. Rev. C} {\bf 104}  (2021)   054320.

\bibitem{D1M}
S.~Goriely, S.~Hilaire, M.~Girod and S.~P\'eru, {\em Phys. Rev. Lett.} {\bf
  102}  (2009)   242501.

\bibitem{Gogny}
{J. Decharge and M. Girod and D. Gogny}, {\em Phys. Lett. B} {\bf 55}  (1975)
  361.

\bibitem{kibedi2002}
T.~Kib\'edi and R.~H. Spear, {\em At. Data and Nucl. Data Tables} {\bf 80}
  (2002) 35 .

\bibitem{DEANGELIS2002}
G.~{de Angelis \it{et al.}}, {\em Phys. Lett. B} {\bf 535}  (2002) 93.

\bibitem{IBFM}
F.~Iachello and P.~{Van Isacker}, {\em The interacting boson-fermion model}
  (Cambridge Univ. Press, Cambridge, UK, 1991).

\bibitem{iachello1979}
F.~Iachello and O.~Scholten, {\em Phys. Rev. Lett.} {\bf 43}  (1979) 679.

\bibitem{nomura2018oct}
K.~Nomura, T.~Nik\ifmmode \check{s}\else \v{s}\fi{}i\ifmmode~\acute{c}\else
  \'{c}\fi{} and D.~Vretenar, {\em Phys. Rev. C} {\bf 97}  (2018)   024317.

\bibitem{scholten1985}
O.~Scholten, {\em Prog. Part. Nucl. Phys.} {\bf 14}  (1985) 189.

\bibitem{rzacaurban2012}
T.~{Rzaca-Urban \it{et al.}}, {\em Phys. Rev. C} {\bf 86}  (2012)   044324.

\bibitem{federman1977}
P.~Federman and S.~Pittel, {\em Phys. Lett. B} {\bf 69}  (1977) 385 .

\bibitem{heyde1987}
K.~{Heyde \it{et al.}}, {\em Nucl. Phys. A} {\bf 466}  (1987) 189 .

\bibitem{bengtsson1987}
R.~{Bengtsson \it{et al.}}, {\em Phys. Lett. B} {\bf 183}  (1987) 1.

\bibitem{nazarewicz1993}
W.~Nazarewicz, {\em Phys. Lett. B} {\bf 305}  (1993) 195 .

\bibitem{duval1981}
P.~D. Duval and B.~R. Barrett, {\em Phys. Lett. B} {\bf 100}  (1981)   223.

\bibitem{garciaramos2014}
J.~E. Garc\'{\i}a-Ramos and K.~Heyde, {\em Phys. Rev. C} {\bf 89}  (2014)
  014306.

\bibitem{gavrielov2019}
N.~Gavrielov, A.~Leviatan and F.~Iachello, {\em Phys. Rev. C} {\bf 99}  (2019)
   064324.

\bibitem{leviatan2018}
A.~{Leviatan \it{et al.}}, {\em Phys. Rev. C} {\bf 98}  (2018)   031302.

\bibitem{nomura2022octcm}
K.~Nomura, {\em Phys. Rev. C} {\bf 106}  (2022)   024330.

\bibitem{Garrett1999-112Cd}
P.~E. {Garrett \it{et al.}}, {\em Phys. Rev. C} {\bf 59}  (1999) 2455.

\bibitem{cd114-2003}
D.~{Bandyopadhyay \it{et al.}}, {\em Phys. Rev. C} {\bf 68}  (2003)   014324.

\bibitem{clement2007}
E.~{Cl\'ement \it{et al.}}, {\em Phys. Rev. C} {\bf 75}  (2007)   054313.

\bibitem{aprahamian2018}
A.~{Aprahamian \it{et al.}}, {\em Phys. Rev. C} {\bf 98}  (2018)   034303.

\bibitem{xiang2020}
J.~{Xiang \it{et al.}}, {\em Phys. Rev. C} {\bf 101}  (2020)   064301.

\bibitem{nomura2020pv}
K.~Nomura, D.~Vretenar, Z.~P. Li and J.~Xiang, {\em Phys. Rev. C} {\bf 102}
  (2020)   054313.

\bibitem{vanisacker1982}
P.~{Van Isacker \it{et al.}}, {\em Nucl. Phys. A} {\bf 380}  (1982) 383 .

\bibitem{zamfir2002}
N.~V. Zamfir, J.-y. Zhang and R.~F. Casten, {\em Phys. Rev. C} {\bf 66}  (2002)
    057303.

\bibitem{sugita1996}
M.~Sugita, T.~Otsuka and P.~von Brentano, {\em Phys. Lett. B} {\bf 389}  (1996)
  642 .

\bibitem{iachello1982}
F.~Iachello and A.~Jackson, {\em Phys. Lett. B} {\bf 108}  (1982) 151.

\bibitem{daley1983}
H.~Daley and F.~Iachello, {\em Phys. Lett. B} {\bf 131}  (1983) 281.

\bibitem{spieker2015}
M.~Spieker, S.~Pascu, A.~Zilges and F.~Iachello, {\em Phys. Rev. Lett.} {\bf
  114}  (2015)   192504.

\bibitem{bijker2017}
R.~Bijker and F.~Iachello, {\em Nucl. Phys. A} {\bf 957}  (2017) 154.

\bibitem{otsuka1986}
T.~Otsuka, {\em Phys. Lett. B} {\bf 182}  (1986) 256 .

\bibitem{otsuka1988}
T.~Otsuka and M.~Sugita, {\em Phys. Lett. B} {\bf 209}  (1988) 140 .

\bibitem{dobaczewski2005}
J.~Dobaczewski and J.~Engel, {\em Phys. Rev. Lett.} {\bf 94}  (2005)   232502.

\end{thebibliography}

\end{document}